# X-ray chemistry in the envelopes around young stellar objects

P. Stäuber[1], S.D. Doty[2], E.F. van Dishoeck[3], and A.O. Benz[1]

[1] Institute of Astronomy, ETH-Zentrum, CH-8092 Zurich, Switzerland
[2] Department of Physics and Astronomy, Denison University, Granville, OH 43023, USA
[3] Sterrewacht Leiden, PO Box 9513, 2300 RA Leiden, The Netherlands

**Abstract.** We present chemical models of the envelope of a young stellar object (YSO) exposed to a central X-ray source. The models are applied to the massive star-forming region AFGL 2591 for different X-ray fluxes. Model results for this region show that the X-ray ionization rate with and without the effects of Compton scattering differs by only a few percent and the influence of Compton scattering on the chemistry is negligible. The total X-ray ionization rate is dominated by the 'secondary' ionization rate of $H_2$ resulting from fast electrons. The abundance profiles of several molecular and atomic species are shown to depend on the X-ray luminosity and on the distance from the source. The carbon, sulphur and nitrogen chemistries are discussed. It is found that $He^+$ and $H_3^+$ are enhanced and trigger a peculiar chemistry. Several molecular X-ray tracers are found and compared to tracers of the far ultraviolet (FUV) field. Like ultraviolet radiation fields, X-rays enhance simple hydrides, ions and radicals. In contrast to ultraviolet photons, X-rays can penetrate deep into the envelope and affect the chemistry even at large distances from the source. Whereas the FUV enhanced species cover a region of $\approx$ 200–300 AU, the region enhanced by X-rays is $\gtrsim$ 1000 AU. We find that $N_2O$, HNO, SO, $SO^+$, $HCO^+$, $CO^+$, $OH^+$, $N_2H^+$, $SH^+$ and $HSO^+$ (among others) are more enhanced by X-rays than by FUV photons even for X-ray luminosities as low as $L_X \approx 10^{30}$ erg s$^{-1}$. $CO_2$ abundances are reduced in the gas-phase through X-ray induced FUV photons. For temperatures $T \lesssim 230$ K, $H_2O$ is destroyed by X-rays with luminosities $L_X \gtrsim 10^{30}$ erg s$^{-1}$. Best-fit models for AFGL 2591 predict an X-ray luminosity $L_X \gtrsim 10^{31}$ erg s$^{-1}$ with a hard X-ray spectrum $T_X \gtrsim 3 \times 10^7$ K. This is the first time that the X-ray flux of a highly obscured source has been estimated by its envelope chemistry. Furthermore, we find $L_X/L_{bol} \approx 10^{-6}$. The chemistry of the bulk of the envelope mass is dominated by cosmic-ray induced reactions rather than by X-ray induced ionization for X-ray luminosities $L_X \lesssim 10^{33}$ erg s$^{-1}$. The calculated line intensities of $HCO^+$ and $HCS^+$ show that high-$J$ lines are more affected than lower $J$ lines by the presence of X-rays due to their higher critical densities, and that such differences are detectable even with large aperture single-dish telescopes. Future instruments such as Herschel-HIFI or SOFIA will be able to observe X-ray enhanced hydrides whereas the sensitivity and spatial resolution of ALMA is well-suited to measure the size and geometry of the region affected by X-rays.

**Key words.** stars: formation – stars: individual: AFGL 2591 – ISM: molecules – X-rays: ISM

## 1. Introduction

Observational studies of star-forming regions show that some young stellar objects (YSOs) are very strong X-ray emitters. Typical X-ray luminosities range from approximately $L_X = 10^{28}$ erg s$^{-1}$ to $L_X = 10^{33}$ erg s$^{-1}$ in the 0.5–10 keV band (e.g., Hofner & Churchwell 1997, Carkner et al. 1998, Feigelson & Montmerle 1999). The heating of the X-ray emitting plasma is not well understood. In low-mass YSOs the emission may originate from powerful magnetic activity near the stellar surface or in the star-disk environment, whereas in high-mass YSOs wind instabilities and shocks may cause the high X-ray flux. In the earliest stage of evolution, the protostar is still deeply embedded in its natal molecular cloud ($A_V > 100$ mag). As a consequence, X-rays are not directly observable toward very young objects, and the onset of the high energy radiation re-

*Send offprint requests to*: P. Stäuber, e-mail: `pascalst@astro.phys.ethz.ch`

mains a secret to this day. Indeed, X-ray observations toward massive star-forming regions are still rare, which may be due to absorption by the large hydrogen column densities toward these objects (Grosso et al. 2005).

Molecular gas exposed to X-rays forms an X-ray dissociation region (XDR) with a peculiar chemistry and physical structure. There has been a growing interest in XDRs in the past twenty years and several models have been developed in order to study these regions. Krolik & Kallman (1983) investigated the influence of X-ray ionization on the Orion molecular cloud assuming a fixed density and temperature. Lepp & McCray (1983) presented constant gas pressure models to calculate the temperature and infrared line emission from an interstellar gas cloud containing a compact X-ray source. Maloney et al. (1996) studied the influence of X-rays on the physical and chemical state of neutral gas over a wide range of densities and X-ray fluxes. In addition, they discussed diagnostic line ratios to distinguish XDRs from shocks and photodissocia-tion regions (PDRs). However, their chemical network focused primarily on carbon and oxygen, whereas nitrogen-bearing species, for example, were neglected. Calculations of molecular abundances for varying X-ray ionization rates in interstellar clouds were presented by Lepp & Dalgarno (1996). They mainly concentrated on HCO$^+$ and nitrogen-containing compounds. A thorough and more general treatment of the physics and chemistry in molecular clouds exposed to X-rays was done by Yan (1997). Tiné et al. (1997) calculated the infrared response of H$_2$ in dense clouds. Other models concentrate on the X-ray ionization of protoplanetary disks (e.g., Glassgold et al. 1997, Aikawa & Herbst 1999, Markwick et al. 2002) or planetary nebulae (e.g., Natta & Hollenbach 1998). Most recently, Meijerink & Spaans (2005) presented a code for photodissociation and X-ray dissociation regions and discussed thermal and chemical differences between the two regions. They calculated four depth-dependent models for different densities and radiation fields that are typical in starburst galaxies and active galactic nuclei.

In this paper we study the influence of a central X-ray source on the chemistry in the envelopes around massive YSOs, using updated atomic and molecular data. The goal of this investigation is to find X-ray tracers that are observable with current or future telescopes in the (sub)millimeter or near-infrared range. The challenge is to distinguish between far ultraviolet (FUV) tracers (Stäuber et al. 2004) and X-ray tracers, since both kinds of high-energy radiation tend to form ions and radicals. Further aims of this study are to estimate not only the X-ray flux emitted by highly obscured objects but also the ionization rate and ionization fraction in the envelopes around YSOs by carefully studying the chemistry. The ionization fraction of a cloud is an important parameter in the formation of a star as magnetic fields control the dynamics of ions. Ambipolar diffusion may support the molecular cloud and regulate the process of mass accretion. It may also influence disk viscosity and jet acceleration. Due to the small atomic and molecular cross sections at high energy (the total cross section summed over all species for a 1 keV photon is $\approx 2.5 \times 10^{-22}$ cm$^2$ and decreases with energy as $\approx \lambda^3$), X-rays can penetrate deeper into the envelope than, for example, FUV photons and affect the gas-phase chemistry even at large distances from the source. In addition, the X-ray ionization rate may exceed the cosmic-ray ionization rate for a large part of the envelope. X-rays are therefore a plausible candidate for the ionization source in the inner, dense part of a YSO envelope.

We have extended the time- and position-dependent chemical model of Doty et al. (2002) to allow the impact of X-rays on the envelope. The physical and chemical models are described in Sect. 2 and Sect. 3. Our results are presented and discussed in Sect. 4. A selection of possible X-ray tracers is discussed in Sect. 5. In Sect. 6 best-fit models for AFGL 2591 are evaluated. To compare the modeled abundance profiles to observations and to show that the influence of X-rays on the chemistry is observable with (large) existing single-dish telescopes, we have calculated emission lines for a selection of species using the Monte Carlo radiative transfer code of Hogerheijde & van der Tak (2000). These results are shown in Sect. 7. Although our calculations are focused on high-mass YSOs, the qualitative results should be equally applicable to the envelopes around low-mass YSOs. Doty et al. (2004) showed that the chemical models of the low-mass source IRAS 16293–2422 required only minor modifications to their high-mass model. We summarize and conclude this paper in Sect. 8.

## 2. Model

The model is based on the detailed thermal and gas-phase chemistry models of Doty et al. (2002) and has been extended to allow the impact of X-rays on the envelope chemistry of young stellar sources.

### 2.1. X-ray flux

The observed X-ray spectra from high-mass YSOs are usually fitted with the emission spectrum of a thermal plasma (e.g., Hofner & Churchwell 1997, Hofner et al. 2002). The thermal X-ray spectrum can be approximated with

$$F_{\rm in}(E,r) = F_0(r)e^{-E/kT_{\rm X}} \quad \left[\text{photons s}^{-1}\,\text{cm}^{-2}\,\text{eV}^{-1}\right], \quad (1)$$

where $F_{\rm in}(E,r)$ is the incident X-ray flux per unit energy, and $T_{\rm X}$ is the temperature of the X-ray emitting plasma. If the X-ray luminosity $L_{\rm X}$ of the source is known, the factor $F_0(r)$ can be evaluated from

$$L_{\rm X} \equiv 4\pi r^2 F_{\rm X} = 4\pi r^2 \int F_{\rm in}(E) E dE \quad \left[\text{ergs s}^{-1}\right]. \quad (2)$$

The column density between the source and our first calculated point is always $N_{\rm H,in} > 1/\sigma_{\rm p}(1\,\text{keV})$, where $\sigma_{\rm p}$ is the total photoabsorption cross section defined in eq. (3). It can therefore be assumed that all photons with energies below 1 keV are absorbed in the inner region. Photons at energies above 100 keV can be neglected as they contribute only minutely to the total integrated X-ray flux. Thus, we calculate the integral (2) over the energy range $E_{\rm min} = 1\,\text{keV}$ to $E_{\rm max} = 100\,\text{keV}$.

The local (i.e., attenuated) X-ray flux per unit energy is finally given by $F(E,r) = F_{\rm in}(E,r)e^{-\tau(E)}$, where $\tau(E)$ is the total X-ray attenuation and $F_{\rm in}$ is the incident X-ray flux given by eq. (1). At lower energies ($E \lesssim 10$ keV), X-rays lose their energy mainly through photoabsorption. Assuming that the photoabsorption cross section of a molecule or atom is equal to its photoionization cross section, the attenuation is given by

$$\tau_{\rm p}(E) = N_{\rm H}\sigma_{\rm p}(E) = N_{\rm H} \sum x(i)\sigma_i(E), \quad (3)$$

where,

$$x(i) = \frac{n(i)}{n({\rm H}) + 2n({\rm H}_2)}, \quad (4)$$

and $N_{\rm H}$ is the total hydrogen column density, $\sigma_{\rm p}$ the total photoabsorption cross section given by the sum of the ionization cross sections of each species in the cloud multiplied by its fractional abundance $x(i)$. We have adopted the analytic fits of Verner et al. (1996) for the photoionization cross sections of atoms and ions and use the cross sections provided by Yan (1997) for H$_2$. To allow for the attenuation of X-rays by grains, the elemental abundances in the solid phase are added to eq.

(3). The assumed total and gas-phase abundances are given in Table A.1. The values listed in Table A.1 are taken from Doty et al. (2002) for the gas-phase abundances and from Yan (1997) for the total abundance.

For higher energies, X-rays also lose energy through inelastic Compton scattering, and Compton ionization becomes the dominant ionization source. The highly energetic X-ray photons interact mainly with free and bound electrons in the gas. Since molecular hydrogen is the most abundant species in our models, the total Compton cross section is dominated by hydrogen, rather than by heavy elements. We have fitted the values provided by the XCOM-NIST database (Berger et al. 1999) for hydrogen and assume that the Compton cross section of molecular hydrogen is twice that of atomic hydrogen. The energy loss of a Compton scattered photon is negligible compared to its initial energy, and attenuation through Compton scattering becomes important only for hydrogen column densities $N_H \gtrsim 10^{24}$ cm$^{-2}$. However, since each scattering process leads to an ionization, the total (effective) cross section for ionizing $H_2$, $\sigma_{eff}$ (see Sect. 3.2), is the sum of the total photoabsorption cross section $\sigma_p$ and the cross section for Compton scattering $\sigma_s$. The attenuation of the X-rays is mainly through photoabsorption. Figure 1 shows the photoabsorption cross sections for the element abundances given in Table A.1, the Compton cross section of molecular hydrogen, and the total (effective) ionization cross section $\sigma_{eff}$.

Model results for AFGL 2591 show that the X-ray ionization rate with and without the effects of Compton scattering differs by at most 20% and that the influence of Compton scattering on the chemistry is negligible. Nevertheless, we include attenuation and ionization due to Compton scattering, using a simplified radiative transfer method to calculate the incident radiation field.

Although eq. (1) approximates a thermal energy distribution, the X-ray spectrum of a thermal plasma has a more complex structure. In addition to the bremsstrahlung emission, line emission can become important. In order to evaluate the accuracy of the assumed X-ray spectrum, we have fitted the results of Raymond & Smith (1977) for an X-ray spectrum of a hot plasma. The results are very similar, leading to the same best-fit models (Sect. 6.1). The results in the following sections are therefore presented using the X-ray spectrum given by eq. (1).

### 2.2. Physical and thermal model

We have applied our model to the massive star-forming region AFGL 2591. AFGL 2591 is taken as a prototypical example of a deeply embedded high-mass YSO (a so-called high-mass protostellar object, HMPO) for which extensive observational data exist and for which reference models without X-rays are available (Doty et al. 2002, Stäuber et al. 2004). No X-ray emission, however, has been observed to date toward this source. Assuming a distance of 1 kpc, the bolometric luminosity is $\sim 2 \times 10^4 L_\odot$ with a total mass of $\sim 10 M_\odot$ (van der Tak et al. 1999).

We adopt the temperature and power-law density distribution proposed by van der Tak et al. (1999) and Doty et al.

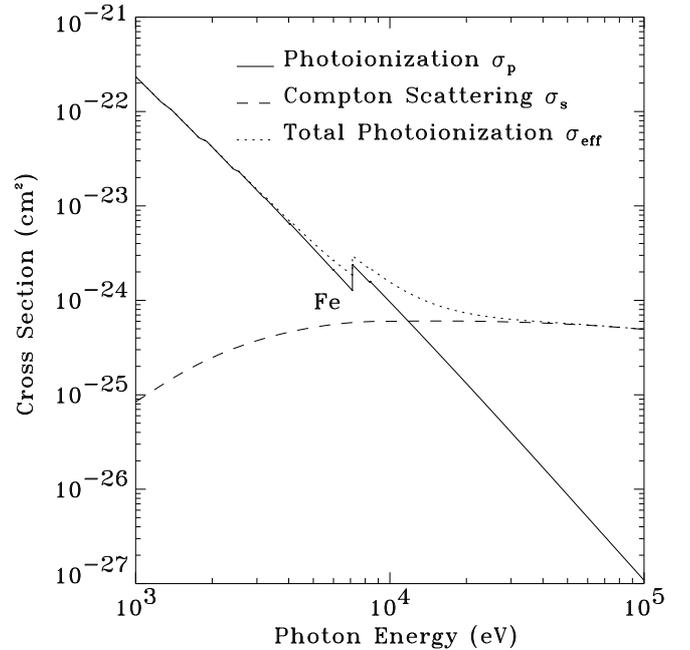

**Fig. 1.** The photoionization cross sections for the elemental abundances given in Table A.1. The cross section for Compton scattering is calculated for hydrogen.

(2002). The density structure of the envelope has been constrained from continuum observations of the dust and CS emission lines over a large range of critical densities. The dust temperature profile was determined from the self-consistent solution of the continuum radiative transfer problem. The gas temperature was calculated explicitly by Doty et al. (2002) who found that $T_{gas} \approx T_{dust}$. The influence of the X-rays on the gas temperature can be estimated by comparing the cooling rate due to gas-dust collisions (Hollenbach & McKee 1989) and the X-ray heating rate provided by Maloney et al. (1996) $n\Gamma_X \approx 3 \times 10^{-1} n H_X$ ergs cm$^{-3}$ s$^{-1}$, where $H_X$ is the energy deposition rate per particle, defined by $H_X = \int F(E,r)\sigma_{eff}(E)E dE$. Our first point of interest is at $r \approx 200$ AU from the central source where the density is already fairly high ($n \approx 10^7$ cm$^{-3}$) but $H_X$ is low due to absorption and geometric dilution ($H_X \approx 10^{-28}$–$10^{-23}$ ergs s$^{-1}$, for X-ray luminosities $L_X \approx 10^{29}$–$10^{33}$ ergs s$^{-1}$). The cooling rate at this point is $\Lambda \approx 3.6 \times 10^{-18} (T_{gas} - T_{dust})$ ergs cm$^{-3}$ s$^{-1}$. Thus, in our modeled regions of AFGL 2591, a meaningful increase in the gas temperature ($T_{gas} - T_{dust} \gtrsim 10$ K) can only be expected for very high X-ray luminosities $L_X \gtrsim 10^{33}$ ergs s$^{-1}$. For our best-fit models (Sect. 6.1), a difference of only 3 K at most is estimated between the gas and dust temperature. We therefore neglect additional heating of the gas through X-rays and assume $T_{gas} \approx T_{dust}$ throughout the envelope. This is a reasonably good approximation as was shown by Doty & Neufeld (1997) and Doty et al. (2002). Figure 2 shows the adopted model for AFGL 2591. The model covers a region from $r_{in} \approx 200$ AU to $r_{out} \approx 29\,000$ AU.

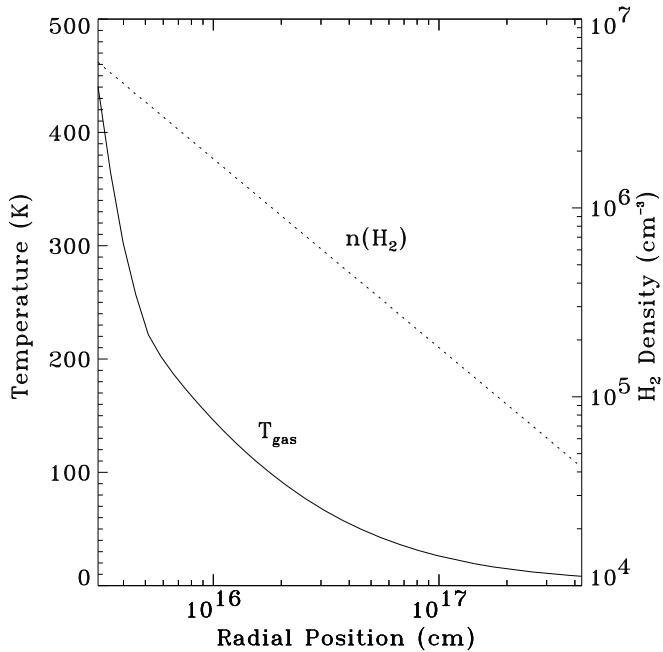

**Fig. 2.** Physical and thermal structure of AFGL 2591. The power-law density distribution is adopted from the model of van der Tak et al. (1999). The gas temperature profile was calculated by Doty et al. (2002).

### 2.3. Chemical model

The chemical model is based upon the UMIST gas-phase chemical reaction network (Millar et al. 1997) and calculates the time-dependent number density $n(i)$ of each species at a certain distance from the source by solving the equations for molecular evolution given by

$$\frac{dn(i)}{dt} = \sum_j k_{ij} n(j) + \sum_{j,k} l_{ijk} n(j) n(k), \quad (5)$$

where $n(i)$ is the number density of species $i$; $k_{ij}$ and $l_{ijk}$ are the rate coefficients for production and destruction of the given species. The main inputs of the model are the temperature and density distribution, the cosmic ray ionization rate and the strength of the far ultraviolet (FUV) radiation field from the inside and outside, $G_{0,in}$ and $G_{0,out}$ respectively. They were discussed in detail by Doty et al. (2002) and Stäuber et al. (2004). The new input parameters concerning the X-rays are the X-ray luminosity $L_X$, the plasma temperature $T_X$ and the inner hydrogen column density $N_{H,in}$, which is the column density between the source and the first innermost grid point of our envelope. In order to study the influence of X-rays on the chemistry, the chemical network has been extended by a number of species and reactions discussed in the following sections.

The assumed initial abundances given in Table A.1 are taken to be consistent with the models of Doty et al. (2002) and Stäuber et al. (2004). The initial gas-phase abundances allow us to reproduce many of the results of the hot core models of Charnley (1997). The effects of freeze-out onto dust grains are included by initially depleting certain species below 100 K. The temperature dependence of molecular depletion is discussed in detail by Boonman et al. (2003a) and Doty et al. (2004). The initial form of sulphur is not known. In our models we assume that sulphur is frozen out onto grains in the form of $H_2S$ and is evaporated into the gas-phase for $T > 100$ K. Solid $H_2S$ has not been detected, however (e.g., Boogert et al. 2000). On the other hand, Wakelam et al. (2004) have shown that it is likely that sulphur is in the icy mantles either as atomic S or in a form, perhaps polymerized S or $S_2$, that is soon converted into S. The dependence of our model results on the initial form of sulphur is discussed in Sect. 4.3.

For the total elemental abundances given in Table A.1 we follow Yan (1997). These abundances are the sum of the gas-phase elemental abundances and the abundances on the dust grains that do not enter the chemistry. In addition to the total hydrogen abundance, the total elemental abundances are important for the attenuation of the X-rays (see Sect. 2.1).

## 3. X-ray induced chemistry

### 3.1. Direct X-ray ionizations and dissociations

X-rays ionize heavy elements preferentially by removing the K-shell electron. The vacancy is then filled by a cascade of radiative (fluorescent) and non-radiative Auger transitions. During this process other electrons and X-ray photons are emitted by the ion, leading finally to a multiply ionized species. The fluorescence probability is less than 10 % for most species and approximately 30 % for Fe (Dwek & Smith 1996). The contribution of the diffuse X-ray emission by fluorescence to the total X-ray flux can therefore be neglected.

We consider the ionization of atoms and atomic ions leading to a singly and doubly ionized state by calculating explicitly the cross sections according to Verner et al. (1996). The probability distribution for the number of ejected electrons for inner shell ionizations are taken from Kaastra & Mewe (1993). The (primary) X-ray ionization rate of species $i$ at a point $r$ is then simply given by

$$\zeta_i = \int F(E,r) \sigma_i(E) dE \quad [s^{-1}], \quad (6)$$

where $\sigma_i(E)$ is the X-ray photoabsorption cross section of species $i$ at energy $E$ and $F(E,r)$ is the local X-ray flux.

Little is known about the impact of X-rays on molecules. Although there are a few cross sections for photoabsorption in the literature, the destruction channels and branching ratios of the dissociated and ionized species are widely unknown. Following Maloney et al. (1996) we therefore consider only diatomic molecules for direct X-ray impact and assume that the molecule dissociates into singly charged ions after inner shell ionization. The cross sections for this process are calculated by adding the atomic cross sections. In general, primary X-ray ionization plays only a minor role in the chemistry since the reactions are $\approx 1000$ times slower compared to the relevant chemical reactions and more than 10 times slower than electron impact ionizations for the case of our AFGL 2591 model parameters.

## 3.2. X-ray induced electron impact reactions

The fast photoelectrons and Auger electrons carry the bulk of the initial X-ray photon energy and are therefore very efficient in ionizing other species. The 'secondary' ionization rate dominates the total ionization rate in XDRs (e.g., Maloney et al. 1996). The electrons can also excite hydrogen and helium. The electronically excited states of H, He and $H_2$ decay back to the ground states by emitting UV photons. The internally generated ultraviolet photons can photoionize and photodissociate other species in the gas – similar to the case of cosmic-ray induced chemistry (e.g., Gredel et al. 1989). Nearly all these secondary processes induced by electron impact are more important for the chemical network than the primary interaction of the X-rays with the gas.

### 3.2.1. Ionizations

Energy deposition of fast electrons in a gas is characterized by a mean energy per ion pair $W(E)$ – the initial energy $E$ of the electron divided by the number of produced ionizations, $N(E)$, in the gas (e.g., Voit 1991, Dalgarno et al. 1999). To calculate $W(E)$ and therefore the number of secondary electrons, we follow Dalgarno et al. (1999) for a H, He and $H_2$ gas mixture. An electron with an initial energy of 1 keV will then lead to $\approx 27$ ionizations. The 'secondary' ionization rate per hydrogen molecule at depth $r$ can be calculated by

$$\zeta_{H_2} = \int F(E,r) \sigma_{\text{eff}}(E) \frac{E}{W(E) x(H_2)} dE \qquad [\text{s}^{-1}], \qquad (7)$$

where $x(H_2) \approx 0.5$ is the fractional abundance of molecular hydrogen and $\sigma_{\text{eff}}$ is the effective or total photoionization cross section at energy $E$, including the cross section for Compton scattering, assuming that each scattering process will ionize a hydrogen molecule.

The electron impact ionization rates $\zeta_i$ of other molecules or atoms can be calculated as a first approximation by multiplying the $H_2$ ionization rate eq. (7) with the ratio of the electron impact cross sections of species $i$ to $H_2$ at a specific energy (Maloney et al. 1996). The average electron energy is taken to be 100 eV. The cross sections for the species were taken from Yan (1997), the NIST database (Kim et al. 2004) or estimated. The ratios for most molecules are between $\approx 2-8$. The error due to the energy dependence of the cross sections or to our estimations are therefore within a factor of a few but certainly $\lesssim 10\times$.

### 3.2.2. UV photodissociations and photoionizations

The excited states in the Lyman-Werner bands of $H_2$ determine the internally generated FUV flux. The number of Ly$\alpha$ photons from H atoms are less important due to the relatively small fractional abundance of atomic hydrogen ($x(H) \lesssim 10^{-4}$). Following Gredel et al. (1989) the photodissociation and photoionization rate of species $i$ is given by

$$R_i = \frac{2x(H_2)\epsilon_{\text{LyW}}\zeta_{H_2}P_{i,\text{LyW}} + x(H)\epsilon_{\text{Ly}\alpha}\zeta_H P_{i,\text{Ly}\alpha}}{1-\omega} \qquad [\text{s}^{-1}], \qquad (8)$$

where $\zeta_{H_2}$ and $\zeta_H$ are the $H_2$ and H ionization rate and $x(H_2)$ and $x(H)$ are the fractional $H_2$ and H abundances, respectively, $\omega$ is the grain albedo ($\omega \approx 0.6$) and $P_{i,\text{LyW}}$ and $P_{i,\text{Ly}\alpha}$ are factors that depend on the atomic or molecular photoabsorption cross sections (Gredel et al. 1989, Sternberg & Dalgarno 1995, Lepp & Dalgarno 1996). In eq. (8), $\epsilon_{\text{LyW}} \sim 1.4$ and $\epsilon_{\text{Ly}\alpha} \approx 1.2$ are the number of excitations per ionization calculated from the values provided by Dalgarno et al. (1999). For CO we follow Maloney et al. (1996) who fitted the self-shielding results of Gredel et al. (1987) and found

$$R_{\text{CO}} = 2.7 x(\text{CO})^{-1/2} \left( \frac{T}{1000\,K} \right)^{1/2} x(H_2) \zeta_T \qquad [\text{s}^{-1}], \qquad (9)$$

where $x(\text{CO})$ is the fractional CO abundance and $\zeta_T$ is the total ionization rate approximated by $\zeta_T \approx \zeta_{H_2} + \zeta_{\text{cr}}$, that is the sum of the $H_2$ ionization rate $\zeta_{H_2}$ and the cosmic-ray ionization rate $\zeta_{\text{cr}}$. The 'primary' X-ray ionization rate $\zeta_i$ can be neglected since it is in general much smaller than $\zeta_{H_2}$.

The He excited $2^1P$ state emits a 19.8 eV photon which ionizes mainly $H_2$, CO or H. The photoionization rate per He atom for these species is (Yan 1997)

$$R_{i,\text{He}} = \zeta_{\text{He}} \frac{x(\text{He})\sigma_i}{x(H_2)\sigma(H_2) + x(H)\sigma(H)} \qquad [\text{s}^{-1}], \qquad (10)$$

where $\zeta_{\text{He}}$ is the He excitation rate and $x(\text{He})$ is the fractional He abundance, $x(H_2)$, $x(H)$, $\sigma(H_2)$ and $\sigma(H)$ are the $H_2$ and H fractional abundances, and the $H_2$ and H photoionization cross sections at $\approx 20$ eV and $\sigma_i$ is the photoionization cross section at $\approx 20$ eV for H, $H_2$ and CO, respectively. The excitation rate of helium can be calculated similar to the $H_2$ ionization rate eq. (7) with the mean energy to excite a helium atom given by Dalgarno et al. (1999).

## 3.3. Other reactions

The electron recombination and charge transfer reactions of the doubly-ionized species are presented in Table B.1 and Table B.2. We consider only C, O, N, S, Fe, Ne and Si in the doubly-ionized state. Other ions in this state are neglected due to their low fractional abundances.

For the recombination of ions on grains we have adopted the treatment of Maloney et al. (1996) for all ions in the model. The grain surface recombination of $HCO^+$ is a critical reaction in dense regions (Aikawa et al. 1999) and important for the abundances of CO-bearing molecules. We follow the calculations of Aikawa et al. (1999) and assume branching ratios of 0.7 for the dissociative recombination and 0.3 for the radiative recombination reaction. However, our models for AFGL 2591 are not very sensitive to these values as long as the ratio dissociative/radiative recombination is $> 1$. For example, no noticeable differences were seen between models with ratios of 0.7/0.3 and 0.85/0.15, respectively. In addition, results of models without any grain recombination reactions of $HCO^+$ have shown no difference for species with fractional abundances $x_i \gtrsim 10^{-10}$ and differences within only $\approx 30\%$ for less abundant species. The recombination of $HCO^+$ in all our models is therefore dominated by electron recombination. Aikawa et al. (1999) further suggested that the dissociative recombination reaction might be more likely for most molecules. We have therefore assumed a branching ratio of 0.7/0.3 for all other molecules as well.

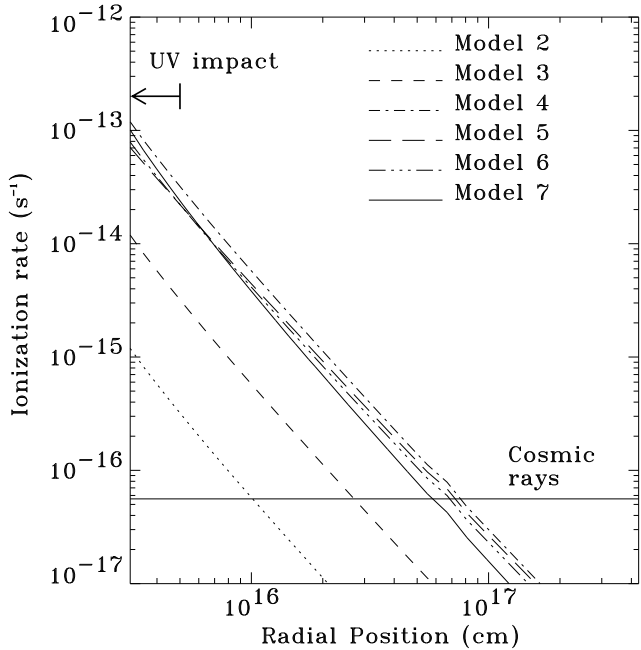

**Fig. 3.** Ionization rates $\zeta_{H_2}$ for the X-ray models presented in Table 1 and a constant cosmic-ray ionization rate $\zeta_{cr} = 5.6 \times 10^{-17}$ s$^{-1}$. The arrow indicates the distance out to which an inner UV field is effective.

**Table 1.** Model parameters for AFGL 2591.

| Model | $L_X$ (ergs s$^{-1}$) | $T_X$ (K) | $N_{H,in}$ (cm$^{-2}$) | $G_{0,in}$ |
|---|---|---|---|---|
| 0 | 0 | 0 | 0 | 0 |
| 1 | 0 | 0 | 3.0E+22 | 10 |
| 2 | 1.0E+30 | 7.0E+07 | 3.0E+22 | 0 |
| 3 | 1.0E+31 | 7.0E+07 | 3.0E+22 | 0 |
| 4 | 1.0E+32 | 7.0E+07 | 3.0E+22 | 0 |
| 5 | 1.0E+32 | 1.0E+08 | 5.0E+22 | 0 |
| 6 | 8.0E+31 | 1.0E+08 | 3.0E+22 | 10 |
| 7 | 6.0E+31 | 3.0E+07 | 3.0E+22 | 10 |

Aside from the grain surface recombination reactions we have ignored grain-surface chemistry, with the exception of the formation of $H_2$. In our model, the rate is $R_{H_2} = 3 \times 10^{-17} n(H) n_H$ cm$^{-3}$ s$^{-1}$, where $n(H)$ is the atomic hydrogen density, and $n_H$ is the density of hydrogen nuclei. Other reactions that affect the $H_2$ network are the formation of the $H^-$ ion by radiative association $H + e^- \rightarrow H^- + h\nu$ and the reaction $H^+ + H \rightarrow H_2^+ + h\nu$. For these rates we have adopted the fits by Latter (1989). Since we are almost always in a pure molecular environment, these reactions play only a minor role in our models though.

## 4. Results and discussion

The chemistry in the envelope around AFGL 2591 under the influence of X-rays has been modeled assuming spherical sym-metry. The parameters that were varied are the X-ray luminosity, the plasma temperature and the inner hydrogen column density. The inner hydrogen column density $N_{H,in}$ is the X-ray absorbing column density between the central source and the first calculated point at $r \approx 200$ AU (see also eq. (3)). Since only the total hydrogen column density is known, $N_{H,in}$ is treated as an unknown parameter.

The equilibrium timescale for X-ray induced chemistry is $t_{eq} \approx \zeta_{H_2}^{-1} \propto r^2/L_X$ yr. Thus, equilibrium may not be reached for most parts of the envelope. The results in this section are presented for $t \approx 5 \times 10^4$ years according to our best-fit models. This result is in good agreement with the chemical age of AFGL 2591 found by Doty et al. (2002). For simplicity, the luminosity was kept constant with time. For all models an outer standard FUV field $G_{0,out} = 1$ has been assumed to be consistent with Doty et al. (2002) and Stäuber et al. (2004). The effects of an enhanced outer FUV radiation field will be discussed in Sect. 4.7.

Table 1 lists the models that are discussed in more detail in this paper. In addition to the best-fit models (Model 5, 6, 7), models with a plasma temperature $T_X = 7 \times 10^7$ K and an inner column density of $N_{H,in} = 3 \times 10^{22}$ cm$^{-2}$ are presented. Model 0 corresponds to the standard model of Doty et al. (2002) without any inner radiation field and Model 1 assumes an inner FUV radiation field with a field strength $G_{0,in} = 10$ according to the models of Stäuber et al. (2004). The plasma temperature of Models 2–4 has been chosen as a mean value of the best-fit models (Sect. 6.1). The only parameter that has been varied for the plots of the Models 2–4 is the X-ray luminosity $L_X$. This is useful since the energy deposition rate $H_X$ is directly proportional to the X-ray luminosity and all the X-ray induced reactions therefore scale with $L_X$. Results for different X-ray temperatures $T_X$ and inner hydrogen column densities $N_{H,in}$ are presented in Sect. 4.6.

Figure 3 shows the $H_2$ ionization rates $\zeta_{H_2}$ for the different X-ray models. In addition, the cosmic-ray ionization rate $\zeta_{cr} = 5.6 \times 10^{-17}$ s$^{-1}$ is shown which is taken from van der Tak & van Dishoeck (2000). By taking the physical parameters given in Fig. 2, it can be estimated that half of the envelope mass is inside $r \approx 3 \times 10^{17}$ cm. Comparison of $\zeta_{H_2}$ and $\zeta_{cr}$ in Fig. 3 therefore shows that the bulk of the envelope mass is dominated by the cosmic-ray ionization rate rather than by the X-ray ionization rate. X-ray luminosities $L_X \gtrsim 10^{33}$ ergs s$^{-1}$ are required for the X-ray ionization rate to dominate the cosmic-ray ionization rate. The influence of different cosmic-ray ionization rates on the chemistry is shown in Sect. 4.8.

Abundances of various species are presented in Fig. 4–6 for Models 0, 2–4 and Model 6. Since it is more common to compare number densities to the number density of molecular hydrogen rather than to the total hydrogen density, the following results are presented with fractional abundances $\tilde{x}(i) = n(i)/n(H_2)$.

### 4.1. CO, $H_2O$ and $CO_2$

The CO abundance profile in Fig. 4 shows that the total abundance of this molecule is not significantly affected by X-rays.

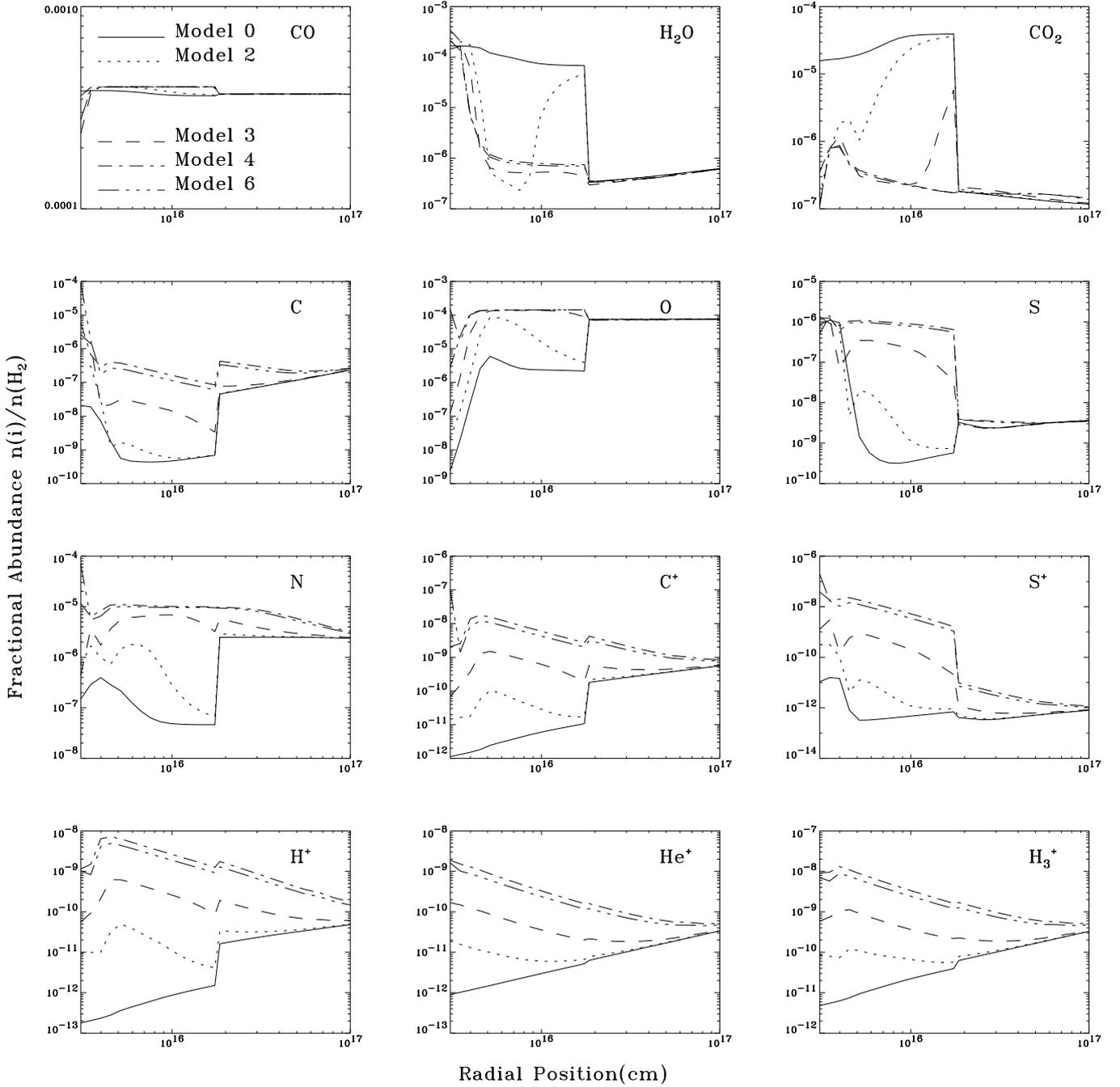

**Fig. 4.** Depth dependent fractional abundances for the models described in Table 1. The solid line corresponds to Model 0, the dotted line is Model 2, the dashed line is Model 3. The dashed-dotted line is Model 4 and the dashed-dotted-dotted line (Model 6) correspond to the best fitted X-ray model with an inner UV field of $G_{0,in} = 10$.

$H_2O$ and $CO_2$ on the other hand are destroyed in the gas-phase. $CO_2$ is destroyed mainly by FUV photons induced through electron impact on $H_2$. The additional $CO_2$ at higher temperatures is primarily from the reaction of OH with CO. The bump at $r \approx 4 \times 10^{15}$ cm ($T \approx 300$ K) is due to the reaction of $HCO_2^+$ with CO which is slightly faster at this distance than the destruction of $CO_2$. $HCO_2^+$ is efficiently produced by the reaction of $HCO^+$ with OH. Detailed studies of Boonman et al. (2003b) showed that the fractional $CO_2$ abundance is $\tilde{x}(CO_2) \approx 1$–$2 \times 10^{-6}$ for $T \gtrsim 300$ K and $\tilde{x}(CO_2) \approx 10^{-8}$ for $T \lesssim 300$ K. This jump at $T \approx 300$ K is well produced by our models for relatively high X-ray luminosities $L_X \gtrsim 10^{31}$ ergs s$^{-1}$. The models of Doty et al. (2002) for AFGL 2591 overpredicted the $CO_2$ abundance in the gas phase and Stäuber et al. (2004) showed that an inner FUV flux does not destroy $CO_2$ significantly either. X-rays are therefore a possible explanation for the observed gas-phase $CO_2$ abundance profile.

$H_2O$ is most efficiently destroyed by reactions with $HCO^+$ and $H_3^+$ for $T \gtrsim 100$ K. In the warm, inner region water is mainly produced by the reaction of OH with $H_2$. For temper-atures $T \lesssim 230$ K, however, most of the gas-phase OH and O goes into $O_2$, thus $H_2O$ is less abundant for these temperatures. $H_2O$ has been observed and studied in detail towards AFGL 2591 by Boonman & van Dishoeck (2003). They derived a fractional abundance of $\tilde{x}(H_2O) \approx 10^{-4}$ for $T \gtrsim 100$ K (scenario 8 in Boonman & van Dishoeck 2003). Our models for high X-ray fluxes ($L_X \gtrsim 10^{30}$ ergs s$^{-1}$) give such high abundances only at $T \gtrsim 230$ K and are therefore inconsistent with the observations of gas-phase water towards AFGL 2591. The low X-ray luminosity is in contradiction with the overall best-fit results (Sect. 6.1) and with the results for $CO_2$. One possible solution to this problem could be the evolution of water with time or related to this, the X-ray luminosity may not be constant with time. However, the destruction of $H_2O$ starts already at $t \gtrsim 1000$ yrs which would imply an implausibly young age. Another possible origin of the discrepancy may be the chemical reaction coefficients that cause the production of $O_2$ rather than $H_2O$ for $100$ K $\lesssim T \lesssim 230$ K. In particular, the coefficients may depend on the populations of the fine-structure levels of atomic oxygen. A higher temperature from $r \approx 0.4$–$1.7 \times 10^{16}$ cm would solve this problem as well.

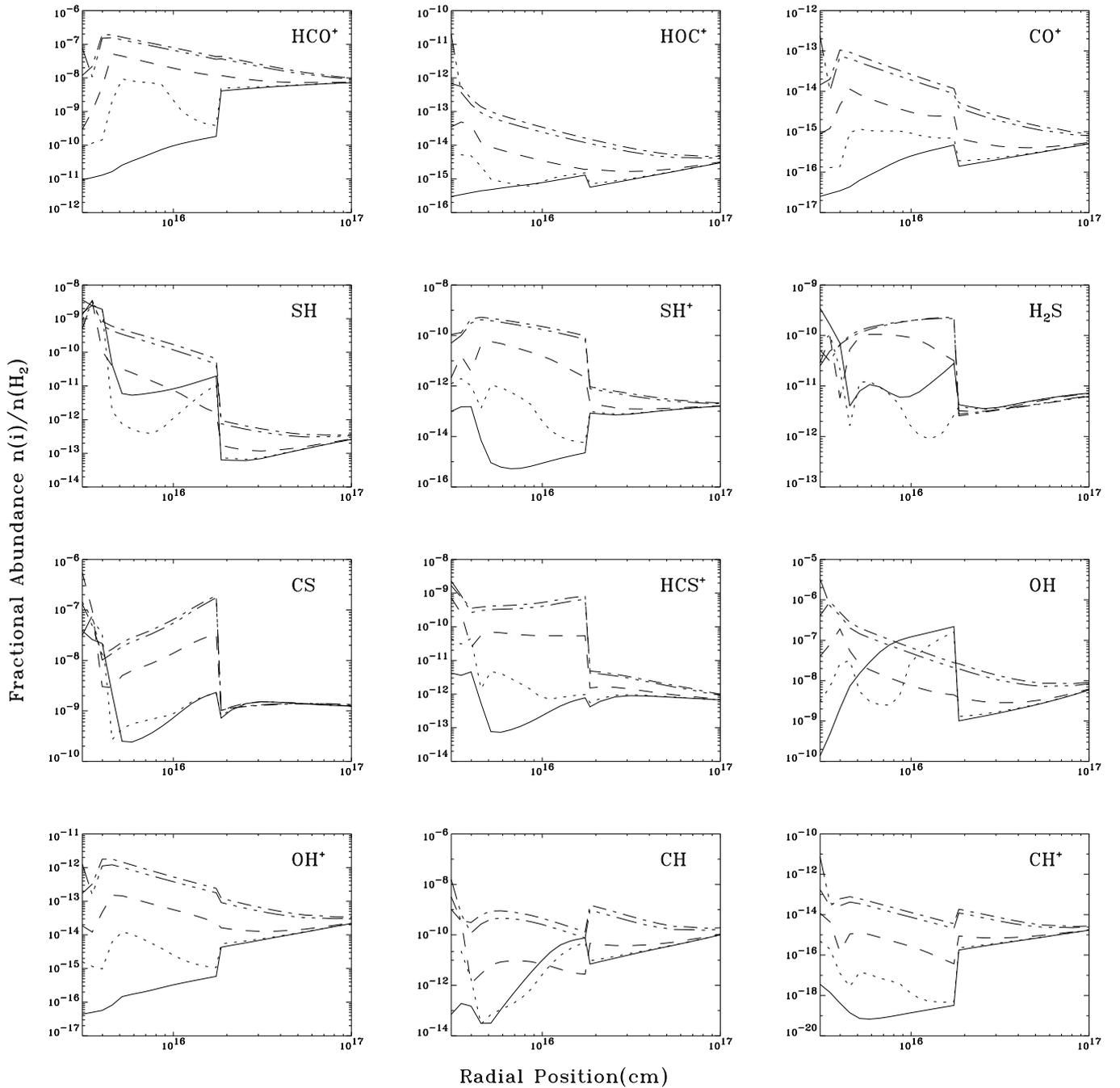

**Fig. 5.** Depth dependent fractional abundances for the models described in Table 1. The solid line corresponds to Model 0, the dotted line is Model 2, the dashed line is Model 3. The dashed-dotted line is Model 4 and the dashed-dotted-dotted line (Model 6) correspond to the best fitted X-ray model with an inner UV field of $G_{0,\mathrm{in}} = 10$.

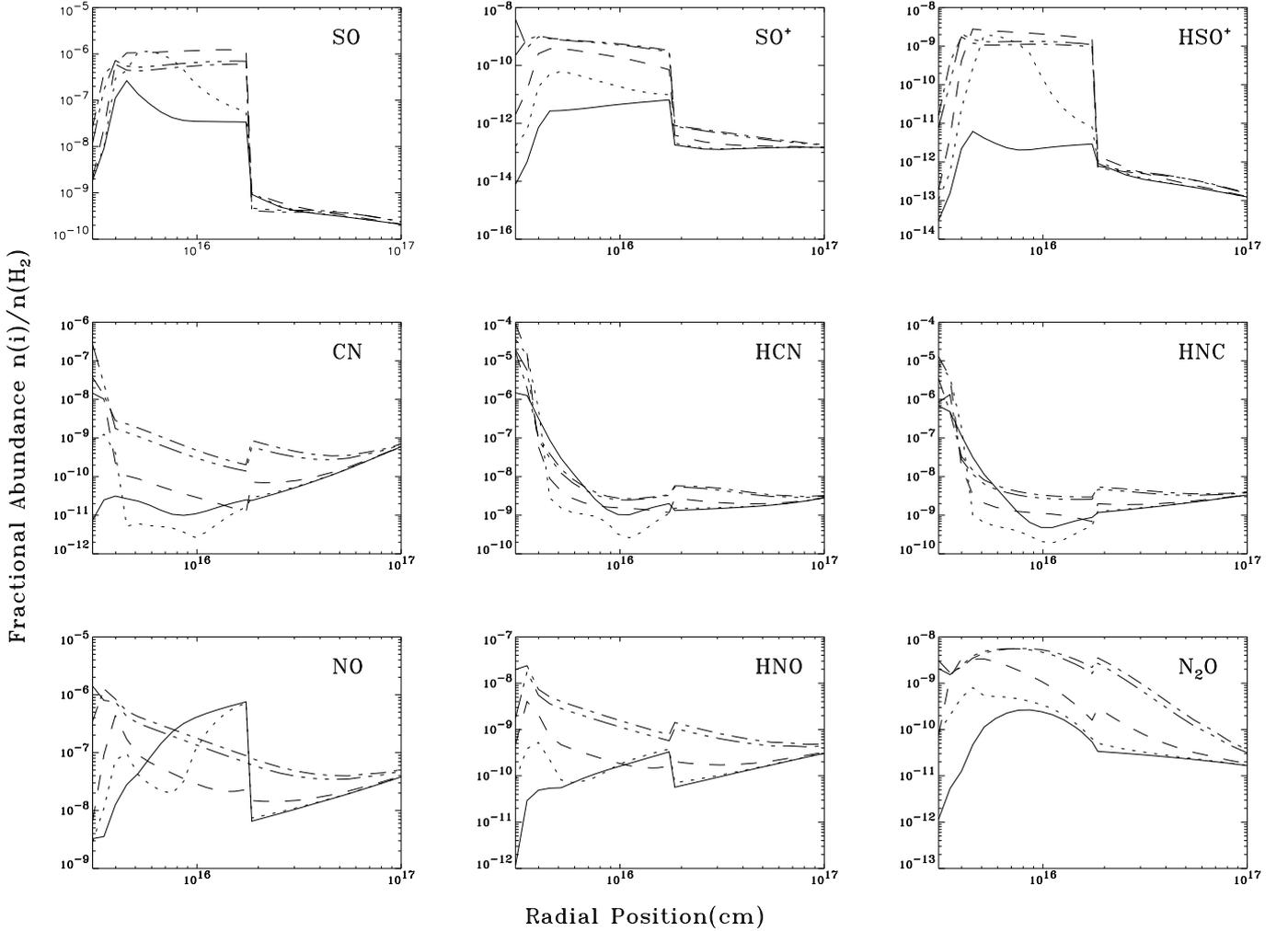

**Fig. 6.** Depth dependent fractional abundances for the models described in Table 1. The solid line corresponds to Model 0, the dotted line is Model 2, the dashed line is Model 3. The dashed-dotted line is Model 4 and the dashed-dotted-dotted line (Model 6) correspond to the best fitted X-ray model with an inner UV field of $G_{0,in} = 10$.

### 4.2. Carbon: C, $C^+$, $CO^+$, $HCO^+$ and $HOC^+$

In the models with X-rays, atomic carbon is particularly produced by the photodissociation of CO by FUV photons from $H_2$ excited by electron impact. If a FUV field $G_{0,in}$ is turned on, however, carbon is more efficiently created by the photons from the inner FUV field. By comparing the two reaction rates, it can be shown that the production of atomic carbon by dissociating CO is dominated by X-ray induced reactions if the total ionization rate is

$$\zeta_T \gtrsim 3 \times 10^{-12} \left(\frac{x(CO)}{10^{-4}}\right)^{1/2} \left(\frac{T}{300\,K}\right)^{-1/2} G_0 e^{-2.5 A_V} \quad [s^{-1}], \quad (11)$$

where $A_V$ is the total visual extinction. If we define the difference of a chemical photodissociation region (PDR) and an X-ray dominated region (XDR) by the way CO is photodissociated, eq. (11) allows to determine the physical boundary of the two regions through the visual extinction $A_V$. For Models 4–7, that is the best-fit models, this is the case for $A_V \approx 2$.

Neutral atomic carbon is mainly destroyed by ionization through FUV photons from either X-ray induced processes or from the inner FUV field. Further in the envelope ($T \lesssim 100\,K$), carbon is primarily destroyed in reactions with $O_2$ and $HCO^+$. The most efficient way to produce $C^+$ in the pure X-ray models is the reaction $He^+ + CO \rightarrow C^+ + O + He$. For high X-ray fluxes ($L_X \gtrsim 5 \times 10^{31}\,erg\,s^{-1}$), this reaction even dominates the photoionization due to an inner FUV field once $A_V \gtrsim 2$. The production of $C^+$ through the X-ray induced FUV photoionization of C or through the 'primary' X-ray ionization and dissociation of CO (Sect. 3.1) is approximately 10 times less efficient. The fastest destruction process for $C^+$ is by reaction with $H_2O$. This is also the dominant production mechanism for $HOC^+$. $HCO^+$ on the other hand, is more efficiently produced by the reaction of $H_3^+$ with CO. This is due to the fact that the production of $HCO^+$ through this reaction is $\approx 60$ times faster compared to the production of $HOC^+$. The destruction of $HCO^+$ and $HOC^+$ is mainly through reactions with water in the inner part of the envelope and through electron recombination at larger distances from the source. HOC$^+$ is also destroyed in reactions with H$_2$ which form HCO$^+$.

CO$^+$ is mainly produced by electron impact ionization of CO. Other important reactions are the charge exchange reaction of He$^+$ with CO$_2$ and the reactions of C$^+$ with OH and O$_2$, respectively. CO$^+$ is quickly destroyed by H$_2$ and forms HCO$^+$.

### 4.3. Sulphur chemistry

Fig. 4–6 show that sulphur and sulphur-bearing molecules are strongly enhanced by X-rays. The following (simple) network explains the relatively high fractional abundances of some of these species: The chemical sulphur network starts with the production of atomic sulphur through X-ray induced FUV photodissociation of H$_2$S. Sulphur reacts then with OH to form either SO or SH. SO reacts with atomic carbon and can therefore form CS. Reactions of CS with H$_3$O$^+$ and HCO$^+$ lead to HCS$^+$. SH$^+$ is efficiently produced by the reactions of atomic sulphur with H$_3^+$ or HCO$^+$. The dominant destruction channel for S in the innermost part of the envelope is the photoionization due to an inner FUV field or due to FUV photons from X-ray induced reactions. These are also the most important reactions to produce S$^+$. S$^+$ reacts with OH and O$_2$ and forms SO$^+$. X-rays can ionize atomic sulphur even at large distances and the abundances of S$^+$, SO$^+$ and HCS$^+$ are enhanced throughout the entire envelope.

In order to study the dependence of our model results on the form of the initial sulphur, we have run models assuming the initial sulphur to be atomic for $T > 100$ K instead of being in the form of H$_2$S with the following result: For the best fitted X-ray models (Sect. 6.1) all species have exactly the same abundances after a few 1000 yrs as in the models where the initial sulphur was assumed to be in H$_2$S. In models without X-rays, however, the results differ less than 20% once $t \gtrsim 5 \times 10^4$ yrs. For $t \approx 3 \times 10^4$ yrs, the chemical age of AFGL 2591 found by Doty et al. (2002), the column density of H$_2$S is $\approx 25$ times less in the model where the initial sulphur was in atomic form compared to the model where sulphur was in the form of H$_2$S. By comparing the two models without X-rays but with different initial forms of sulphur to observations of H$_2$S (van der Tak et al. 2003) it is seen that the model where sulphur is assumed to be initially in the form of H$_2$S fits the observations much better. Other species like S, SO$_2$, CS and H$_2$CS, however, differ less than 30% for $t \approx 3 \times 10^4$ yrs. Since we will compare our results to the models of Doty et al. (2002) and Stäuber et al. (2004) we assume sulphur initially to be in the form of H$_2$S.

### 4.4. Nitrogen chemistry

Since the initial nitrogen in our models is in molecular form, the nitrogen chemistry starts with the photodissociation of N$_2$ by either an inner FUV field $G_{0,\text{in}}$ or by X-ray induced FUV photons from excited H$_2$. Atomic nitrogen is then transformed into NO in reactions with OH. NO reacts with N and forms N$_2$, or it reacts with C to make CN. HCN is then efficiently produced by the reaction of CN with H$_2$. The destruction of N$_2$ by H$_3^+$ leads to N$_2$H$^+$ that is mainly destroyed by H$_2$O or by CO. The destruction of N$_2$H$^+$ by HCN leads to HNCH$^+$ which can recombine either back to HCN, HNC, or CN.

The bulk abundance of N$^+$ in the gas stems from the ionization of N$_2$ by He$^+$. For N$^+$, reactions with H$_2$ lead to NH$_4^+$ which dissociates into NH$_2$ and H$_2$ via electron recombination. NH$_2$ and O form HNO which on the other hand forms N$_2$O by reaction with O. From Fig. 6 it can be seen that both HNO and N$_2$O are greatly enhanced by X-rays.

### 4.5. Simple Hydrides

Hydrides are strongly enhanced by X-rays as can be seen in Fig. 4–6. The hydrides SH and SH$^+$ have already been discussed in Sect. 4.3. X-ray enhanced CH reaches a fractional abundance of $\tilde{x}(\text{CH}) \approx$ few $\times 10^{-10}$ for high X-ray fluxes ($L_X \approx 10^{32}$ ergs s$^{-1}$) and its abundance is almost constant in the entire envelope. CH is mainly produced by the reaction CH$_2$ + H $\rightarrow$ CH + H$_2$. CH$_2$ is efficiently produced in reactions of C with H$_2$ and by the X-ray induced photodissociation of CH$_4$. At larger distances of the source, electron recombination reactions of CH$_3^+$ become also important for the production of both CH and CH$_2$. CH is primarily destroyed in reactions with H$_2$. CH$^+$ is mainly formed in reactions of H$_3^+$ or HCO$^+$ with atomic carbon. The fractional abundance of CH$^+$ is also almost constant throughout the envelope for the higher X-ray flux models, but with $\tilde{x}(\text{CH}^+) \approx 10^{-15}$–$10^{-13}$, $\tilde{x}(\text{CH}^+) < 10^{-3} \times \tilde{x}(\text{CH})$. However, CH$^+$ is enhanced up to $\tilde{x}(\text{CH}^+) \approx 10^{-13}$ in the innermost part of the envelope where it is $\approx 10^5$ times more abundant in the best-fit models than in the models without X-rays.

OH and OH$^+$ are both enhanced by X-rays through the destruction of water in the gas-phase. OH is the product of FUV photodissociation of H$_2$O whereas OH$^+$ is mainly produced in reactions of He$^+$ with H$_2$O or in reactions of H$_3^+$ with OH. Both, OH and OH$^+$ are destroyed primarily by H$_2$. For the best-fit models, OH reaches fractional abundances $\tilde{x}(\text{OH}) \approx 10^{-8}$ in the outer part and $\tilde{x}(\text{OH}) \approx 10^{-6}$ in the innermost part of the envelope. OH$^+$ has a fractional abundance of $\tilde{x}(\text{OH}^+) \approx 10^{-14}$–$10^{-13}$ in the outer part and reaches $\tilde{x}(\text{OH}^+) \approx 10^{-12}$ closer to the source.

Hydrides like SH, SH$^+$, CH, CH$^+$, OH and OH$^+$ are difficult to observe with ground based telescopes due to atmospheric absorption. These hydrides are therefore possible molecules to observe with the space-borne instrument Herschel-HIFI (de Graauw & Helmich 2001).

### 4.6. Varying $T_X$ and $N_{\text{H,in}}$

The influence of different X-ray temperatures $T_X$ and inner hydrogen column densities $N_{\text{H,in}}$ on the chemistry is shown in Fig. C.1 for six species for $L_X = 10^{32}$ ergs s$^{-1}$. It can be seen that the results for $T_X = 3 \times 10^7$ K and $T_X = 7 \times 10^7$ K are very similar. A lower temperature of $T_X = 7 \times 10^6$ K, however, leads to lower abundances at larger distances from the source since the softer X-rays are more absorbed on their way through the envelope. The results for different inner column densities are even less distinctive. In general, higher absorbing hydrogen column densities lead to lower abundances. Comparison of the integrated column densities for all species for the different models shows that the results differ only a factor of 2 at most for the different X-ray temperatures, $T_X$, and absorbing inner hydrogen column densities, $N_{H,in}$. The influence of the X-ray temperature and the inner hydrogen column density on the chemistry is therefore not as strong as it is for the X-ray luminosity where the abundances of species like $H_3^+$ basically scale with the X-ray luminosity (see also Table 2 in Sect. 5). Indeed, it will be shown in Sect. 6.1 that the X-ray luminosity $L_X$ is the dominant parameter in our models.

### 4.7. Influence of an enhanced outer FUV field

Regions of high-mass star formation often have FUV fields that are higher than the average interstellar radiation field. Evolved massive stars emit a copious amount of ultraviolet radiation. The immediate environment is ionized and a H II region is formed. The FUV photons (6< $h\nu$ < 13.6 eV), however, penetrate the H II region and may influence the chemistry of a nearby cloud or the envelope of a young stellar object, forming a photon-dominated region (eg., Hollenbach & Tielens 1999). For AFGL 2591, a nearby H II region has been found (van der Tak et al. 1999, Trinidad et al. 2003) whose exciting star may have an influence on the outer part of the AFGL 2591 envelope.

Fig. D.1 shows a selection of species for different outer FUV fields $G_{0,out}$. The influence of the outer FUV field is noticeable at $r \gtrsim 2.5 \times 10^{17}$ cm from the source. The chemistry is that of a photon-dominated region (e.g., Sternberg & Dalgarno 1995). The chemical reaction network basically starts with the dissociation of CO and the ionization of C leading to an enhancement of atomic and ionized carbon. Other species that are enhanced by an outer FUV field are CH and $CH^+$. In our models, CH is mainly produced by the dissociation of $CH_4$ and $CH_2$. Photoionization of CH leads to $CH^+$. $CO^+$ is enhanced through reactions of $C^+$ with OH and $CH^+$ with O. $HOC^+$ is only enhanced for a low outer FUV field ($G_{0,out} = 10$). $HOC^+$ is produced in reactions of water with $C^+$.

Other molecules like $HCO^+$, CN, HCN, $HCS^+$ or $SH^+$ that are enhanced by X-rays are destroyed by an outer FUV field. The increase of the electron abundance leads to relatively fast electron recombination reactions for $HCO^+$, $HCS^+$ and $SH^+$. CN and HCN are efficiently photodissociated by the FUV field. Due to the lower density in the outer part of the envelope, however, these species differ less than 30% in the radial column density for the different outer FUV fields. C, $C^+$, CH and $CH^+$ on the other hand can be enhanced up to several orders of magnitude. Since the influence of the outer FUV field is limited to the outermost part of the envelope, spatially resolved data can distinguish between an abundance enhancement of these species due to an outer FUV field and due to X-rays or FUV photons from the inside. In addition, the excitation of these species is different in the inner part of the envelope where the density is higher than in the outer part of the envelope.

### 4.8. Influence of cosmic-ray ionization rate

Like X-rays, cosmic rays are able to produce $He^+$ and $H_3^+$ and trigger a peculiar chemistry (e.g., Lepp et al. 1987, Herbst 2000). The cosmic-ray ionization rate $\zeta_{cr} = 5.6 \times 10^{-17}$ s$^{-1}$ in our models was derived from observations of $HCO^+$ by van der Tak & van Dishoeck (2000). In order to study the influence of different cosmic-ray ionization rates on our X-ray models we have run Model 5 with a $\approx$ 5 times higher and lower cosmic-ray ionization rate, respectively. Figure D.2 presents the result of this investigation for a few species. Comparison of Model 0 with $\zeta_{cr} = 5.6 \times 10^{-17}$ s$^{-1}$ and Model 5 with $\zeta_{cr} = 10^{-17}$ s$^{-1}$ and $\zeta_{cr} = 5.6 \times 10^{-17}$ s$^{-1}$ shows that the influence of the cosmic-ray ionization rate dominates the chemistry for $r \gtrsim 8 \times 10^{16}$ cm from the source for $L_X = 10^{32}$ ergs s$^{-1}$ in agreement with Fig. 3. Cosmic-ray ionization rates as high as $\zeta_{cr} = 3 \times 10^{-16}$ s$^{-1}$, however, are able to dominate the gas-phase chemistry for $r \gtrsim 2 \times 10^{16}$ cm from the source, that is from where $H_2O$ freezes out on dust grains. By comparing $H^{13}CO^+$ observations to a simple chemical model with different cosmic-ray ionization rates van der Tak & van Dishoeck (2000) showed for seven high-mass sources that the ionization rate varies between $\zeta_{cr} = 0.61$–$5.6\times10^{-17}$ s$^{-1}$. In addition, ionization rates as high as $\zeta_{cr} \approx 10^{-16}$ s$^{-1}$ are not confirmed by our best-fit models (Sect. 6.1). The influence of the cosmic-ray ionization rate on the emission line profiles of $H^{13}CO^+$ will be discussed in Sect. 7 (see also Doty et al. 2002).

## 5. X-ray tracers

In order to find species that trace X-rays in highly obscured regions we have to distinguish X-ray enhanced species from species that are enhanced by other mechanisms such as inner FUV fields from the source, outer FUV fields from nearby stars or from cosmic-ray induced chemistry. By focusing only on the innermost region of the envelope ($r \lesssim 2000$ AU), the effects of the latter two mechanisms can be neglected. The only challenge then is to separate species that are enhanced by X-rays from species that trace more preferably the FUV field from the central source.

### 5.1. X-ray vs. FUV tracers

Figures 4–6 show that X-rays are capable of influencing the atomic and molecular abundances throughout almost the entire envelope. This is the main difference with the models of Stäuber et al. (2004) where the influence of an inner FUV field has been examined. Due to their small cross sections, X-rays penetrate deeper into the envelope and affect the chemistry on large scales. The enhanced region due to a low inner FUV field ($G_{0,in} = 10$) is $\approx 300$–500 AU (Stäuber et al. 2004) whereas the X-ray enhanced region is often $\gtrsim 1000$ AU, even for relatively low X-ray fluxes ($L_X \approx 10^{30}$ ergs s$^{-1}$). Current and future (sub)millimeter interferometers like ALMA are able to resolve regions covering $\lesssim 1$–2″. At a distance of 1 kpc, this corresponds to a region of $\approx 1000$–2000 AU. However, it will be shown in Sect. 7 that the influence of X-rays and therefore the

**Table 2.** AFGL 2591 predicted column densities (cm$^{-2}$) for different X-ray models.

| Species | Model 2 $N_{\rm beam}$ | $q_{2/0}$ | $q_{2/1}$ | Model 3 $N_{\rm beam}$ | $q_{3/0}$ | $q_{3/1}$ | Model 4 $N_{\rm beam}$ | $q_{4/0}$ | $q_{4/1}$ | Model 6 $N_{\rm beam}$ |
|---|---|---|---|---|---|---|---|---|---|---|
| H | 3.1E+16 | 1.9 | 0.1 | 1.7E+17 | 10.6 | 0.6 | 1.6E+18 | 96.5 | 5.2 | 1.4E+18 |
| C | 7.5E+13 | 5.4 | 2.0E-02 | 5.2E+14 | 37.2 | 0.1 | 3.2E+15 | 228.9 | 0.7 | 6.0E+15 |
| N | 8.3E+15 | 7.0 | 2.0 | 1.0E+17 | 84.9 | 24.8 | 1.7E+17 | 145.4 | 42.4 | 1.7E+17 |
| O | 3.6E+17 | 8.4 | 7.2 | 2.2E+18 | 50.2 | 43.6 | 2.4E+18 | 55.8 | 48.4 | 2.4E+18 |
| S | 4.1E+14 | 1.0 | 1.1 | 3.2E+15 | 7.9 | 8.2 | 1.5E+16 | 36.2 | 37.6 | 1.3E+16 |
| CH | 2.0E+11 | 0.4 | 0.2 | 1.9E+11 | 0.4 | 0.2 | 7.4E+12 | 16.8 | 6.5 | 4.7E+12 |
| NH | 4.0E+12 | 0.5 | 0.5 | 2.2E+12 | 0.3 | 0.3 | 7.4E+12 | 0.9 | 0.9 | 4.8E+12 |
| NH$_2$ | 1.6E+13 | 0.9 | 1.1 | 1.2E+13 | 0.7 | 0.8 | 4.0E+13 | 2.2 | 2.7 | 2.4E+13 |
| OH | 8.8E+14 | 0.4 | 0.4 | 2.0E+14 | 0.1 | 0.1 | 1.4E+15 | 0.7 | 0.6 | 1.2E+15 |
| HCN | 3.5E+15 | 9.3 | 0.8 | 6.0E+15 | 15.9 | 1.4 | 1.1E+15 | 2.9 | 0.2 | 1.7E+15 |
| CN | 3.9E+11 | 1.5 | 3.0E-02 | 2.5E+12 | 9.5 | 0.2 | 1.3E+13 | 50.0 | 1.1 | 2.1E+13 |
| HNC | 8.9E+14 | 6.0 | 7.3 | 1.1E+15 | 7.2 | 8.7 | 2.8E+14 | 1.9 | 2.3 | 2.7E+14 |
| N$_2$O | 5.1E+12 | 1.7 | 1.7 | 1.9E+13 | 6.3 | 6.1 | 7.5E+13 | 24.8 | 24.2 | 6.7E+13 |
| NO | 4.4E+15 | 0.6 | 0.6 | 6.8E+14 | 0.1 | 0.1 | 3.4E+15 | 0.5 | 0.5 | 2.7E+15 |
| HNO | 3.7E+12 | 1.2 | 1.2 | 4.5E+12 | 1.5 | 1.5 | 3.3E+13 | 10.9 | 10.9 | 2.4E+13 |
| SH | 8.3E+11 | 0.8 | 0.5 | 6.0E+11 | 0.6 | 0.4 | 3.6E+12 | 3.3 | 2.2 | 2.7E+12 |
| H$_2$S | 7.0E+10 | 0.3 | 0.4 | 1.3E+12 | 5.8 | 6.8 | 3.3E+12 | 14.9 | 17.5 | 3.3E+12 |
| SO | 5.3E+15 | 6.7 | 6.6 | 2.0E+16 | 25.3 | 24.7 | 9.4E+15 | 12.0 | 11.7 | 1.1E+16 |
| CS | 4.2E+13 | 1.5 | 0.8 | 3.2E+14 | 11.6 | 6.2 | 1.4E+15 | 51.1 | 27.1 | 1.2E+15 |
| H$^+$ | 2.2E+11 | 14.1 | 14.1 | 4.4E+12 | 278.9 | 278.2 | 4.7E+13 | 2995.3 | 2988.0 | 3.4E+13 |
| He$^+$ | 1.1E+11 | 2.1 | 2.1 | 6.8E+11 | 12.5 | 12.5 | 6.4E+12 | 117.7 | 117.7 | 4.6E+12 |
| C$^+$ | 5.5E+11 | 5.0 | 0.1 | 1.0E+13 | 93.4 | 2.5 | 1.1E+14 | 982.8 | 26.3 | 8.0E+13 |
| N$^+$ | 1.4E+07 | 2.0 | 2.0 | 7.9E+07 | 11.3 | 11.3 | 7.3E+08 | 103.4 | 103.5 | 5.2E+08 |
| O$^+$ | 1.7E+07 | 7.0 | 2.2 | 1.6E+08 | 65.8 | 20.8 | 1.5E+09 | 631.7 | 199.2 | 1.1E+09 |
| S$^+$ | 1.2E+11 | 8.2 | 1.0E-02 | 4.2E+12 | 294.1 | 0.5 | 1.1E+14 | 7454.9 | 13.1 | 7.8E+13 |
| H$_3^+$ | 1.2E+12 | 2.9 | 2.9 | 6.9E+12 | 17.0 | 17.0 | 6.4E+13 | 158.4 | 158.4 | 4.6E+13 |
| H$_3$O$^+$ | 4.0E+13 | 1.2 | 1.1 | 3.6E+13 | 1.0 | 1.0 | 9.2E+13 | 2.7 | 2.5 | 7.9E+13 |
| OH$^+$ | 5.3E+07 | 8.9 | 3.5 | 9.8E+08 | 164.2 | 65.2 | 9.8E+09 | 1648.7 | 654.7 | 7.0E+09 |
| CH$^+$ | 9.0E+04 | 28.3 | 4.0E-04 | 6.3E+06 | 1973.4 | 3.0E-04 | 2.9E+08 | 9.0E+04 | 1.2 | 4.9E+08 |
| CO$^+$ | 1.4E+07 | 3.1 | 1.1 | 5.9E+07 | 13.0 | 4.4 | 5.0E+08 | 109.3 | 36.7 | 3.7E+08 |
| HCO$^+$ | 4.5E+13 | 25.8 | 9.9 | 3.4E+14 | 193.0 | 74.0 | 1.4E+15 | 783.3 | 300.5 | 1.2E+15 |
| HOC$^+$ | 1.9E+07 | 1.4 | 2.0E-02 | 7.9E+07 | 5.5 | 0.1 | 7.8E+08 | 54.5 | 0.9 | 1.4E+09 |
| HCNH$^+$ | 4.3E+12 | 5.6 | 1.2 | 9.7E+12 | 12.8 | 2.8 | 8.4E+12 | 11.1 | 2.4 | 9.7E+12 |
| N$_2$H$^+$ | 3.7E+11 | 3.8 | 3.8 | 2.2E+12 | 22.7 | 22.7 | 1.8E+13 | 189.9 | 190.1 | 1.3E+13 |
| HCS$^+$ | 3.7E+10 | 5.1 | 0.4 | 1.0E+12 | 135.1 | 10.9 | 1.0E+13 | 1349.3 | 108.7 | 7.8E+12 |
| SO$^+$ | 3.5E+11 | 4.4 | 1.4 | 3.1E+12 | 38.9 | 12.6 | 9.2E+12 | 116.4 | 37.8 | 8.6E+12 |
| HSO$^+$ | 6.4E+12 | 145.2 | 143.2 | 3.4E+13 | 766.1 | 755.6 | 1.9E+13 | 423.9 | 418.1 | 2.2E+13 |
| SH$^+$ | 3.0E+09 | 38.1 | 2.4 | 2.7E+11 | 3494.5 | 218.6 | 3.9E+12 | 5.0E+04 | 3104.7 | 3.0E+12 |

The beam averaged column densities $N_{\rm beam}$ are calculated for a $2''$ beam ($\approx$ 2000 AU). The enhancement factor $q_{x/y}$ is defined by $N_{i,{\rm Model}\,x}/N_{i,{\rm Model}\,y}$. Model 0 is the model without any inner radiation field, Model 1 has no X-rays but an inner FUV field ($G_{0,{\rm in}} = 10$). Model 2–4 are X-ray models with $\log(L_{\rm X}) = 29$, 30 and 31, respectively. Model 6 is one of the best fitted models with an inner FUV field $G_{0,{\rm in}} = 10$, $L_{\rm X} = 8 \times 10^{31}$ ergs s$^{-1}$, $T_{\rm X} = 10^8$ K and $N_{\rm H,in} = 3 \times 10^{22}$ cm$^{-2}$.

enhanced emission of several species can also be seen in much bigger single-dish beams.

X-rays and UV photons have similar effects on the chemistry. They both increase the abundances of ions and radicals. However, X-rays can ionize atoms with an ionization threshold $E_{\rm th} \geq 13.6$ eV like hydrogen, helium, oxygen and nitrogen which is not possible for FUV photons ($6 < h\nu < 13.6$ eV) and can even doubly ionize many species. Reactions with ions like He$^+$ or H$_3^+$ can therefore become important in X-ray induced chemistry. For example, reactions of He$^+$ or H$_3^+$ with CO efficiently produce C$^+$ and HCO$^+$, respectively (Sect. 4.2). The main contribution to the He$^+$ abundance, however, is from electron impact ionization of He, rather than direct X-ray ionization. H$_3^+$ is formed by the reaction of H$_2^+$ with H$_2$ where H$_2^+$ is also formed through electron impact ionization. Both He$^+$ and H$_2^+$ cannot be produced with FUV photons but with cosmic rays as discussed in Sect. 4.8.

Table 2 presents beam averaged column densities of Model 2, 3, 4 and Model 6 for a selection of species for the inner $\approx$ 2000 AU, that is the region that can be resolved by interferometers. The enhancement factor $q_{{\rm Model}\,x/{\rm Model}\,0}$ in Table 2 is defined as the ratio of the column density of a species predicted by the X-ray model $x$ and the column density of the model without any radiation field, that is Model 0. The enhancement factor $q_{\mathrm{Model\,x/Model\,1}}$ is defined as the ratio of the calculated column density from X-ray model $x$ and the column density of the model without X-rays but with an inner FUV field with a field strength $G_{0,\mathrm{in}} = 10$ (Model 1). Since the results for the best-fit models and Model 4 are similar, the column densities of Model 4 are representative of all best fitted models.

Most species presented in Table 2 are enhanced due to X-rays ($q_{\mathrm{Model\,x/Model\,0}} > 1$) and may therefore serve as possible tracers. An enhancement factor $q_{\mathrm{Model\,x/Model\,1}} > 1$, however, distinguishes X-ray tracers from species that are enhanced by an inner UV field. It can be seen in Table 2 that species like $N_2O$, $HNO$, $SO$, $SO^+$, $HCO^+$, $CO^+$, $OH^+$, $N_2H^+$, $SH^+$ and $HSO^+$ are more enhanced by X-rays than by FUV photons even for X-ray luminosities as low as $L_X \approx 10^{30}$ ergs s$^{-1}$. In addition, the abundances for these species increase with increasing X-ray flux. Other species like $CS$, $SH$, $HCN$, $CH$ or $HCS^+$ are enhanced either by X-rays or an inner FUV field, depending on the incident X-ray or FUV flux, respectively. Species like $C$ or $HOC^+$ are more enhanced by an inner FUV field than by X-rays.

Table 2 and Fig. 6 show that $HSO^+$ is a remarkable tracer for X-rays. It is enhanced by more than a factor of 100 even for low X-ray fluxes with fractional abundance $\tilde{x}(HSO^+) \approx 10^{-10}$–$10^{-9}$ in the inner $\approx 1000$ AU. In our models, $HSO^+$ is primarily produced by the reactions of $H_3^+$ and $HCO^+$ with $SO$. Unfortunately though, little is known about this molecule which has not been observed to date. In addition, spectral information is missing and laboratory work is needed before $HSO^+$ can be searched in molecular clouds. $SH^+$ is enhanced even more than 10000 times compared to Model 0 and more than 1000 times compared to Model 1 for the best-fit models. Four hyperfine lines of $SH^+$ have recently been measured by Savage et al. (2004). It was found that the best candidates for observations are the strongest lines near 526 GHz which requires instruments such as Herschel-HIFI or SOFIA.

### 5.2. Column density ratios

Ratios like CN/HCN or $HOC^+/HCO^+$ are of specific interest since they depend on the flux of FUV radiation (Fuente et al. 1995, 2003, Stäuber et al. 2004). Since X-rays affect the abundances of ions and radicals, these ratios may also depend on the incident X-ray flux. The ratios CN/HCN, HNC/HCN, $CO^+/HCO^+$ and $HOC^+/HCO^+$ of the radial column densities of the inner $\approx 2000$ AU are presented in Fig. 7. The ratios $CO^+/HCO^+$ and $HOC^+/HCO^+$ are effectively enhanced only by models containing an inner FUV field. The ratio HNC/HCN is higher for the X-ray models than for the FUV models but similar to the ratio of Model 0. The CN/HCN ratios depend strongly on the incident X-ray flux and vary within a factor of 100. Models with high X-ray luminosities ($L_X \approx 10^{32}$ ergs s$^{-1}$) and models with X-rays and an inner FUV field, that is Model 6 and 7, are capable of increasing the CN/HCN ratio up to 0.01. Although the CN/HCN ratio increases with increasing X-ray flux (Models 2–4), the ratio is also enhanced by an inner FUV field without X-rays. In general, all these ratios give poor information about the X-ray flux, since they are enhanced more likely by an inner FUV flux rather than by X-rays. The crucial point for an enhanced CN/HCN ratio is the strength of the inner FUV field and therefore the inner column density $N_{\mathrm{H,in}}$. For column densities $N_{\mathrm{H,in}} \gtrsim 4 \times 10^{22}$ cm$^{-2}$, the optical depth $A_V$ is $\approx 20$ and most FUV photons from the young star will be absorbed. The enhancement of the CN/HCN ratio is then more likely due to X-rays.

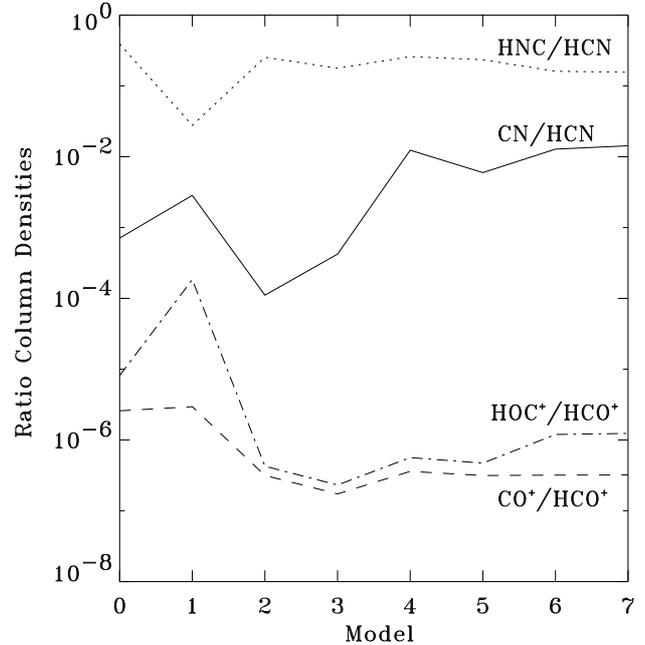

**Fig. 7.** Column density ratios for the inner $\approx 2000$ AU ($\approx 2''$) of the envelope. The models are described in Table 1.

### 5.3. Previously suggested X-ray tracers

Earlier models of molecules exposed to an X-ray source predicted similar molecular X-ray tracers. Lepp & Dalgarno (1996) found that CN and NO are particularly enhanced by X-rays. The enhancement of CN is confirmed with our models. CN can be enhanced up to $\approx 50$ times in the inner 2000 AU. The beam averaged column density of NO given in Table 2, however, is smaller for the X-ray models than for the models without X-rays. From Fig. 6 it can be seen that NO is enhanced in the outer parts of the envelope and in the innermost part of the envelope ($r \lesssim 500$ AU). The 'cut-off' of the bump in the middle part of the envelope causes the decreasing NO column density with increasing X-ray flux. The bump of the NO abundance profile is due to the evaporation of water at $T = 100$ K, a process not included by Lepp & Dalgarno (1996). For $T \gtrsim 100$ K, NO is efficiently produced by the reaction $HNO^+ + H_2O \rightarrow H_3O^+ + NO$. In our X-ray models NO is enhanced up to fractional abundances $\tilde{x}(NO) \approx 10^{-7}$–$10^{-6}$ in the innermost part of the envelope. This is the same result found by Lepp & Dalgarno (1996). To observe the enhanced region of NO towards the inner $\approx 500$ AU of AFGL 2591 the sensitivity and spatial resolution of ALMA is needed. The model results of Krolik & Kallman (1983) for the Orion molecular cloud are generally in good agreement with our results. The enhancement factors they derive are comparable to those of Model 6. In particular, they also find that NO is not changing in models with high ionization rates.

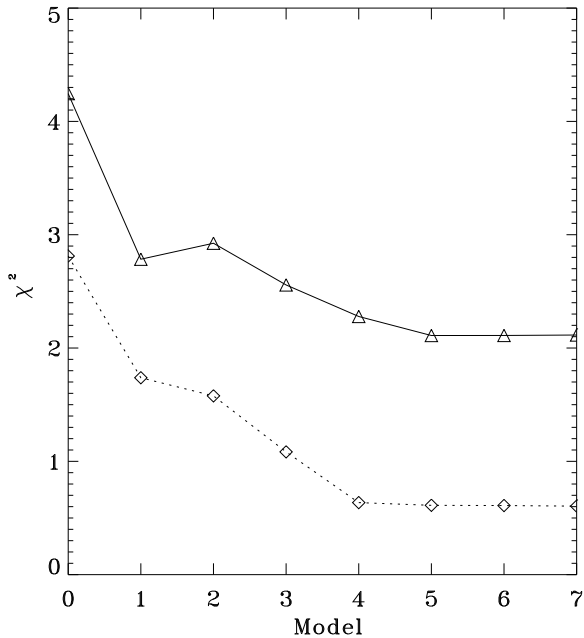

**Fig. 8.** $\chi^2$ values for the Models 0–7 described in Table 1. The upper curve (solid line) corresponds to the $\chi^2$ values calculated for all species, whereas the lower curve (dotted line) describes the values obtained considering only species that are influenced most by X-rays (see text). The $\chi^2$ values are therefore smaller for the dotted curve.

## 6. AFGL 2591 models

### 6.1. Best-fit models

To find the best fitting model for AFGL 2591 by varying the input parameters $L_X$, $T_X$, $N_{H,in}$ – the column density between the X-ray source and the first radial model point – and the inner UV field $G_{0,in}$, a $\chi^2$ statistic has been used defined by

$$\chi^2 = \sum_i w_i^2 \times \frac{(\log(N_{\text{model},i}) - \log(N_{\text{obs},i}))^2}{(\varepsilon_i \times \log(N_{\text{model},i}))^2}. \quad (12)$$

We follow the method and observational data provided by Doty et al. (2002). The weights $w_i$ are in favor of column densities that have been derived by a full non-LTE radiative transfer model ($w_i = 3$) rather than methods assuming LTE ($w_i = 2$). For the error we assume $\varepsilon_i = 0.6$ for all species except for those given either by an upper or a lower limit. For these, we assume $\varepsilon_i = 5$ when the model results are within the limit, since these observational results can have big errors due to, for example, non-detections. The ranking of the fits is not very sensitive to the assumptions on $\varepsilon_i$, though. Agreement of models to within a factor of a few compared to the observations can still be considered as good for chemical modeling. Thus, the ratios of the column densities are taken by calculating the difference of the modeled and observed logarithmic column densities. The value of $\chi^2$ has then a slightly different meaning, but the minimum of eq. (12) still provides the best-fit model since the ideal model would have $\chi^2 = 0$.

As a first approach models with only X-rays are considered: $L_X$ was varied from $10^{29}$ ergs s$^{-1}$ to $7 \times 10^{32}$ ergs s$^{-1}$, $T_X$ from $5 \times 10^6$ K to $3 \times 10^8$ K and $N_{H,in}$ from $10^{22}$ cm$^{-2}$ to $10^{24}$ cm$^{-2}$. The minimum and maximum values of the grid for the X-ray luminosity and plasma temperature are based on results of X-ray observations toward YSOs. The best fitted parameters out of $\approx 800$ calculated models are $L_X = 10^{32}$ ergs s$^{-1}$, $T_X = 10^8$ K, $N_{H,in} = 5 \times 10^{22}$ cm$^{-2}$ and $t \approx 5 \times 10^4$ yrs (Model 5). The X-ray luminosity and temperature are rather high and at the upper end of what has been observed toward star forming regions (e.g., Feigelson & Montmerle 1999, Hofner et al. 2002). If, however, the X-ray luminosity decreases as the YSO evolves – like it does in low-mass YSOs (Feigelson & Montmerle 1999, Tsujimoto 2002) – such high luminosities and plasma temperatures could be possible. The inner column density corresponds to $A_V \approx 25$ which is in good agreement with the results of van der Tak et al. (1999) for $r < 175$ AU.

As a second step in finding the best-fit X-ray model, an inner FUV field of $G_{0,in} = 10$ has been assumed with an inner column density of $N_{H,in} = 3 \times 10^{22}$ cm$^{-2}$ in agreement with the models of Stäuber et al. (2004). The parameters of the best fitted model are then $L_X = 8 \times 10^{31}$ ergs s$^{-1}$, $T_X = 10^8$ K and $t \approx 5 \times 10^4$ yrs (Model 6). Models with higher inner FUV fields in general lead to much higher $\chi^2$ values. The assumption of an inner FUV therefore yields a slightly lower X-ray luminosity.

Using a reduced $\chi^2$ statistic where each addend in eq. (12) is divided by the number of observed lines of that species, the best fitted model is $L_X = 6 \times 10^{31}$ ergs s$^{-1}$, $T_X = 3 \times 10^7$ K with an inner column density of $N_{H,in} = 3 \times 10^{22}$ cm$^{-2}$ and an inner FUV field $G_{0,in} = 10$ at $t \approx 3 \times 10^4$ yrs (Model 7).

Considering only species that are influenced most by X-rays in eq. (12), the best fitted parameters are $L_X = 10^{32}$ ergs s$^{-1}$, $T_X = 9 \times 10^7$ K and $t \approx 4 \times 10^4$ yrs which is very similar to Model 5. The species considered are HCN, HNC, $H_2S$, CS, CN, SO, HCO$^+$, HCS$^+$, $H_3^+$ and $N_2H^+$. The lower (dotted) curve in Fig. 8 shows the results for this calculation whereas the upper (solid) curve corresponds to the $\chi^2$ values for all species.

In general all best-fit models have very similar $\chi^2$ values and the differences may be too small to elect the winner. The goodness of fit of the different parameters is further discussed in the next section. Table E.1 lists the column densities of the observed species and the modeled column densities for Model 6. Further shown are the parameters that fit each specific molecule best. It can be seen that the parameters $L_X$, $T_X$, $N_{H,in}$ and time vary from species to species. Only two species, $H_2CO$ and $C_2H_2$, are modeled best with Model 0, that is the model without any high energy radiation from the source. All other species seem to require a more or less strong X-ray flux from the source to match the observed column density.

O- and B-type main sequence stars emit X-rays due to shocks from the radiation driven stellar winds with $L_X/L_{bol} \approx 10^{-7}$ (Berghöfer et al. 1997). A bolometric luminosity $L_{bol} = 2 \times 10^4$ L$_\odot$ for AFGL 2591 implies an X-ray luminosity $L_X \approx 8 \times 10^{30}$ ergs s$^{-1}$. If we assume that young massive stars emit rather more X-rays, this value could be regarded as a lower limit for the X-ray emission for AFGL 2591. The ratio of the luminosities for our best-fit models is $L_X/L_{bol} \approx 10^{-6}$.

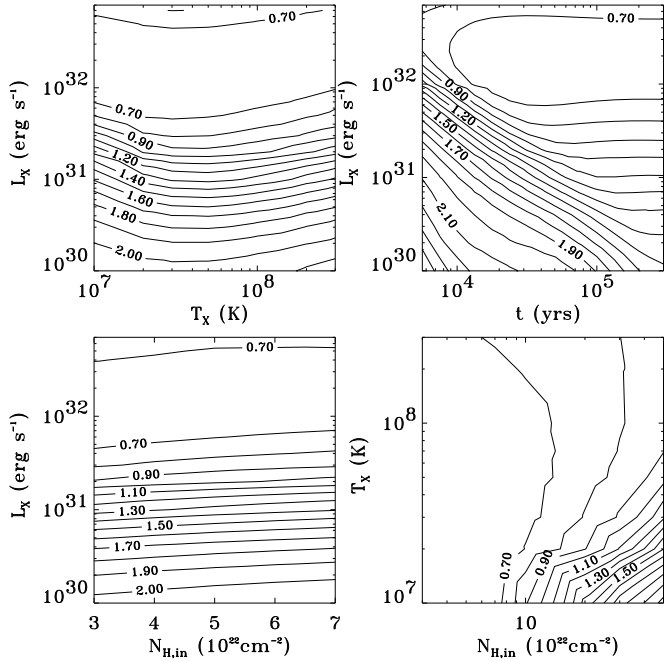

**Fig. 9.** $\chi^2$ contour plots for the species influenced most by X-rays. The parameters in each plot that were not varied have been fixed at $L_X = 10^{32}$ ergs s$^{-1}$, $T_X = 10^8$ K, $N_{H,in} = 5 \times 10^{22}$ cm$^{-2}$ and $t \approx 5 \times 10^4$ yrs (Model 5). The contour levels are plotted for 0.7–2.4 in steps of 0.1.

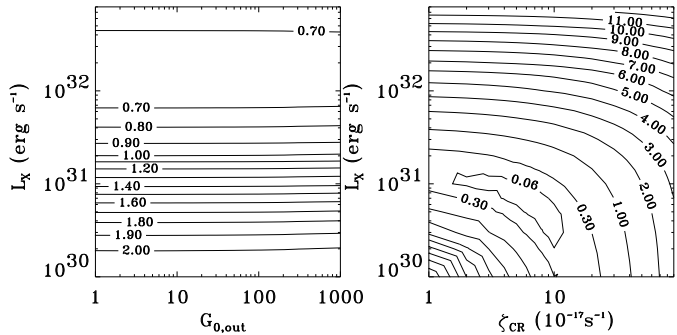

**Fig. 10.** $\chi^2$ contour plot of $L_X$ vs. $G_{0,out}$ considering only species that are most sensitive to X-rays (left) and contour plot $L_X$ vs. $\zeta_{cr}$ considering $\chi^2$ values of HCO$^+$ only (right). The other parameters have been fixed at $T_X = 10^8$ K and $N_{H,in} = 5 \times 10^{22}$ cm$^{-2}$.

As can be seen in Fig. 3, the H$_2$ ionization rates $\zeta_{H_2}$ for the best-fit models (5, 6, 7) differ only little throughout the envelope. The ionization rates of Model 5 and Model 6 are somewhat flatter in radial dependence than that for Model 7 due to the harder spectrum. However, since similar ionization rates lead to similar abundance profiles, the conclusion can be drawn that the X-ray induced chemistry depends mainly on the H$_2$ ionization rate $\zeta_{H_2}$.

### 6.2. Contour plots

To quantify the plausible ranges of the various parameters, contour plots of the $\chi^2$ values for the species influenced most by X-rays are shown in Fig. 9. It can be seen that the results are basically independent of $T_X$ and $N_{H,in}$ for different $L_X$. The plot of $L_X$ vs. $t$ shows that the results become independent of $t$ for high X-ray luminosities once $t \gtrsim 10^4$ yrs. $\chi^2$ does not change with time anymore, thus equilibrium is reached for the species in our statistical sample.

A region of confidence can be defined by the contours that are twice the minimum $\chi^2$ value, which is $\approx 0.6$. The constraint on the X-ray luminosity is then $L_X \gtrsim 10^{31}$ ergs s$^{-1}$ with $t \gtrsim 10^4$ yrs. The plot of $T_X$ vs. $N_{H,in}$ in Fig. 9 shows that for low temperatures ($T_X \approx 10^7$ K), the column density is $N_{H,in} \lesssim 2 \times 10^{23}$ cm$^{-2}$. Higher temperatures allow higher column densities. However, the extra photons at higher energies are not efficiently absorbed by the inner column and the results become independent of the inner column density $N_{H,in}$.

Fig. 10 shows contour plots for $L_X$ vs. $G_{0,out}$ and $L_X$ vs. $\zeta_{cr}$. The plot for the different outer FUV fields shows that the results depend mainly on the X-ray luminosity $L_X$ rather than on the outer FUV field $G_{0,out}$. Models with higher FUV fields than the standard radiation field ($G_{0,out} = 1$) generally lead to higher $\chi^2$ values.

The values in the plot for the cosmic-ray ionization rate have been scaled by a factor of 100 in eq. (12) since only HCO$^+$ was considered. The cosmic-ray ionization rate is well constrained between $\zeta_{cr} \approx 2 \times 10^{-17}$ s$^{-1}$ and $\zeta_{cr} \approx 10^{-16}$ s$^{-1}$ in good agreement with the results of van der Tak & van Dishoeck (2000) and Doty et al. (2002). The X-ray luminosity is well constrained between $L_X \approx 2 \times 10^{30}$ ergs s$^{-1}$ and $L_X \approx 10^{31}$ ergs s$^{-1}$.

## 7. Calculated emission lines

Molecules with high critical densities $n_{crit}$ are likely to probe the inner dense part of the envelope (see, e.g., van Dishoeck & Hogerheijde 1999 for a review). Since the critical density of a molecule is proportional to the Einstein $A$ coefficient for spontaneous transitions and therefore $n_{crit} \propto \nu^3$, higher frequency transitions have also higher critical densities. Stäuber et al. (2004) show that enhancements of molecules due to an inner FUV field should already be detectable with single-dish telescopes having beam widths $\gtrsim 11''$. To show this for the case of X-ray enhanced molecules, the density profiles of HCO$^+$ and HCS$^+$ presented in Sect. 4 are used to compute the line intensities with the Monte Carlo radiative transfer code Hogerheijde & van der Tak (2000). The intrinsic (turbulent) line profile is taken to be a Gaussian with a Doppler parameter of 1.6 km s$^{-1}$, independent of radius. Molecular data are taken from the Leiden atomic and molecular database (Flower 1999, Schöier et al. 2005). The line profiles are convolved to an appropriate telescope beam size. In order to avoid optical depth effects we have modeled the isotope H$^{13}$CO$^+$. The assumed isotope ratio is HCO$^+$/H$^{13}$CO$^+$ = 60 (Wilson & Rood 1994).

Figures 11 and 12 show that high-$J$ transitions are indeed sensitive to the different models and may therefore be used as tracers for high energy radiation. The enhancement factors for the higher transitions presented in Table 3 confirm the results of Sect. 4. H$^{13}$CO$^+$ and HCS$^+$ are clearly enhanced by X-rays. The enhancement factors for the H$^{13}$CO$^+$ $J = 8$–7 transition are between 2–43 compared to Model 0 and between 1–11 compared to the FUV Model 1. The $J = 10$–9 line of $HCS^+$ is enhanced between 2 and 169 times compared to the models without X-rays or FUV field and between 0.2 and 21 times compared to the FUV model. No influence is seen, however, from an enhanced outer FUV field in the emission lines of $H^{13}CO^+$ and $HCS^+$.

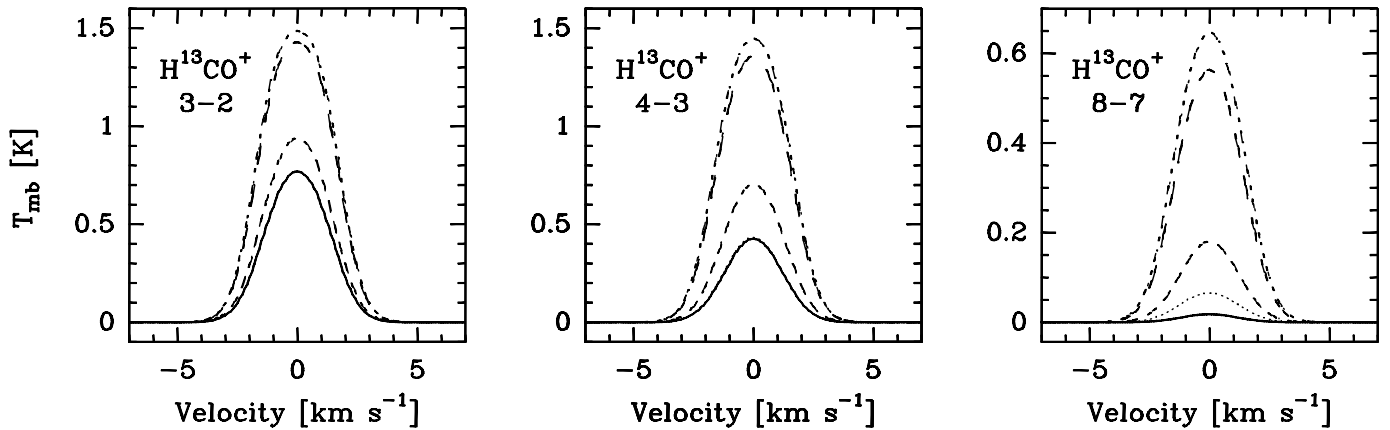

**Fig. 11.** Calculated $H^{13}CO^+$ line profiles for Model 0 (solid line), Model 1 (dotted line), Model 3 (short dashed line), Model 4 (dashed-dotted line) and Model 5 (long dashed line). The $H^{13}CO^+$ 3–2 line (260 GHz) was convolved with a 18″ JCMT beam, the $H^{13}CO^+$ 4–3 line (347 GHz) was convolved with a 14″ JCMT beam and the $H^{13}CO^+$ 8–7 line (694 GHz) was convolved with a 8″ JCMT beam.

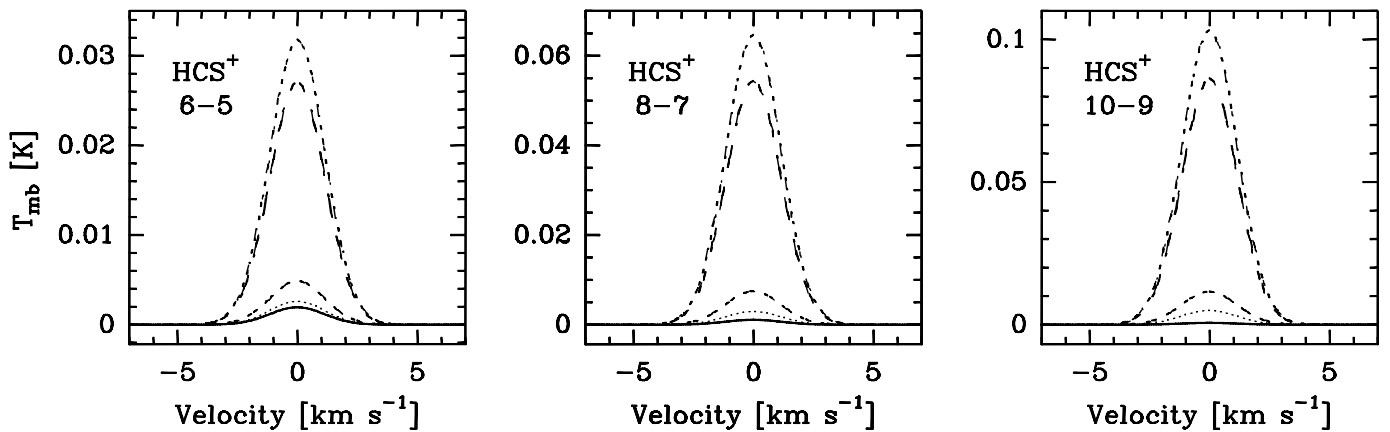

**Fig. 12.** Calculated $HCS^+$ line profiles for Model 0 (solid line), Model 1 (dotted line), Model 3 (short dashed line), Model 4 (dashed-dotted line) and Model 5 (long dashed line). The $HCS^+$ 6–5 line (256 GHz) was convolved with a 18″ JCMT beam, the $HCS^+$ 8–7 line (341 GHz) was convolved with a 14″ JCMT beam and the $HCS^+$ 10–9 line (427 GHz) was convolved with a 11″ JCMT beam.

Figure 13 shows calculated $H^{13}CO^+$ lines for Model 0 and Model 5 with cosmic-ray ionization rates $\zeta_{cr} = 5.6 \times 10^{-17}$ s$^{-1}$ and $\zeta_{cr} = 3 \times 10^{-16}$ s$^{-1}$. As expected, the strongest line is that of Model 5 with $\zeta_{cr} = 3 \times 10^{-16}$ s$^{-1}$. For the $J = 3$–2 and the $J = 4$–3 transitions Model 5 and Model 0 with $\zeta_{cr} = 3 \times 10^{-16}$ s$^{-1}$ differ less than 30–50% and are therefore hardly distinguishable. At higher frequencies, however, the two lines of Model 5 clearly dominate the two lines of Model 0. The $J = 8$–7 transitions of Model 5 are enhanced by a factor of $\approx 5$ compared to that of Model 0 with the higher cosmic-ray ionization rate. Higher $J$ transitions of $H^{13}CO^+$ can therefore be used to distinguish between the effects of a central X-ray source and the effects of an enhanced cosmic-ray ionization. The X-ray luminosity, however, should be of the order of $\approx 10^{31}$ ergs s$^{-1}$ as can be seen in Fig. 11 to distinguish cosmic-ray ionization rates as high as $\zeta_{cr} = 3 \times 10^{-16}$ s$^{-1}$ from the effects of X-rays.

## 8. Conclusion

We have extended the chemical models of Doty et al. (2002) to study the impact of X-rays from a central source on the chemistry of YSO envelopes. The models are applied to the massive star-forming region AFGL 2591 using the physical structure proposed by van der Tak et al. (1999) and Doty et al. (2002). Our major results are summarized below:

1. X-rays can penetrate deep into the envelope due to their small cross sections and affect the chemistry even at large distances from the source. The abundances of many species are enhanced by 2–3 orders of magnitude. Like FUV radiation, X-rays preferably enhance simple hydrides, ions and radicals. The region which is influenced most by X-rays is ≳ 1000 AU. Some species are enhanced throughout the entire envelope. The greater penetration of X-rays is the main difference to the models of Stäuber et al. (2004) where the influence of an inner FUV field on YSO envelopes was studied (Sect. 4).

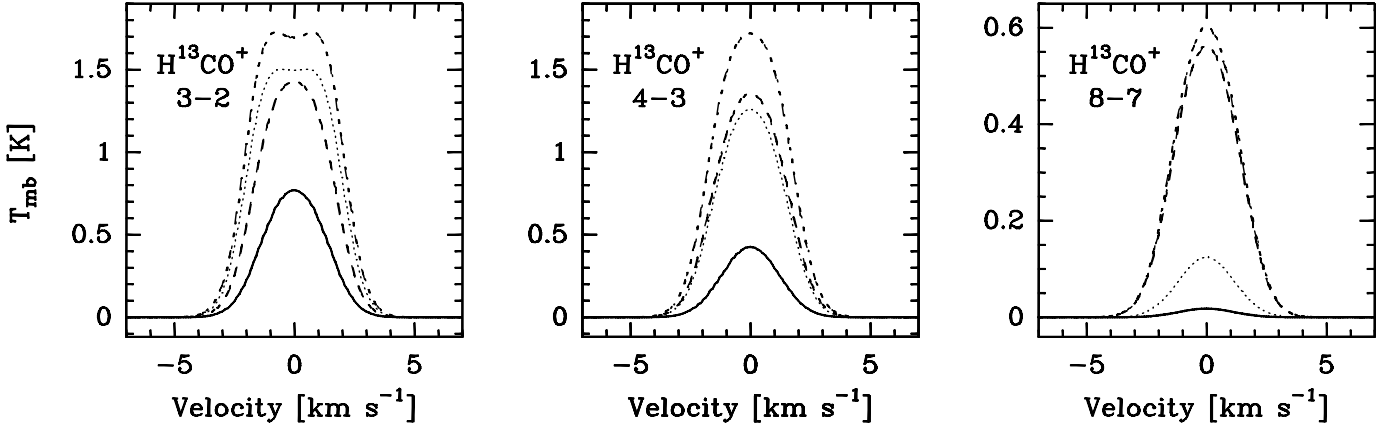

**Fig. 13.** Calculated $H^{13}CO^+$ line profiles for Model 0 (solid line) and Model 5 (dashed line) with a cosmic-ray ionization rate $\zeta_{cr} = 5.6 \times 10^{-17}$ s$^{-1}$ and for Model 0 (dotted line) and Model 5 (dashed-dotted line) with a cosmic-ray ionization rate $\zeta_{cr} = 3 \times 10^{-16}$ s$^{-1}$. The $H^{13}CO^+$ 3–2 line (260 GHz) was convolved with a 18″ JCMT beam, the $H^{13}CO^+$ 4–3 line (347 GHz) was convolved with a 14″ JCMT beam and the $H^{13}CO^+$ 8–7 line (694 GHz) was convolved with a 8″ JCMT beam.

**Table 3.** Enhancement factors $q_{\text{Model x/Model y}} = f_{i,\text{Model x}}/f_{i,\text{Model y}}$ for different line transitions.

| Ratio | $H^{13}CO^+$ 3–2 | $H^{13}CO^+$ 4–3 | $H^{13}CO^+$ 8–7 |
|---|---|---|---|
| $q_{1/0}$ | 1 | 1 | 4 |
| $q_{2/0}$ | 1 | 1 | 2 |
| $q_{3/0}$ | 1 | 2 | 10 |
| $q_{4/0}$ | 2 | 4 | 43 |
| $q_{5/0}$ | 2 | 4 | 36 |
| $q_{6/0}$ | 2 | 4 | 38 |
| $q_{2/1}$ | 1 | 1 | 1 |
| $q_{3/1}$ | 1 | 2 | 3 |
| $q_{4/1}$ | 2 | 4 | 11 |
| $q_{5/1}$ | 2 | 4 | 9 |
| $q_{6/1}$ | 2 | 4 | 10 |
| Ratio | HCS$^+$ 6–5 | HCS$^+$ 8–7 | HCS$^+$ 10–9 |
| $q_{1/0}$ | 1 | 3 | 8 |
| $q_{2/0}$ | 1 | 1 | 2 |
| $q_{3/0}$ | 3 | 7 | 19 |
| $q_{4/0}$ | 17 | 61 | 169 |
| $q_{5/0}$ | 14 | 51 | 141 |
| $q_{6/0}$ | 13 | 49 | 138 |
| $q_{2/1}$ | 1 | 0.5 | 0.2 |
| $q_{3/1}$ | 2 | 2 | 2 |
| $q_{4/1}$ | 12 | 21 | 20 |
| $q_{5/1}$ | 10 | 18 | 17 |
| $q_{6/1}$ | 10 | 17 | 16 |

The values represent the ratios of the integrated line fluxes $f = \int T_{mb} dV$. The Models are described in Table 1 in Sect. 4.

2. He$^+$ and H$_3^+$ have enhanced abundances up to a factor of 1000 in the innermost region compared to models without X-rays. These two ions trigger an X-ray characteristic chemistry which is similar to that induced by cosmic rays. Although the X-ray ionization rate exceeds the cosmic-ray ionization rate for a large part of the envelope, it is found that in the case of AFGL 2591 the chemistry of the bulk of the envelope mass is dominated by the cosmic-ray induced reactions rather than X-ray induced ionization (Sect. 4).

3. The X-ray ionization rate is dominated by the 'secondary' H$_2$ ionization rate resulting from fast electrons. The X-ray induced chemistry therefore depends mainly on the H$_2$ ionization rate $\zeta_{H_2}$ which is directly proportional to the X-ray luminosity (Sect. 4, Sect. 6.1).

4. Several molecules – among them N$_2$O, HNO, SO, SO$^+$, HCO$^+$, CO$^+$, OH$^+$, N$_2$H$^+$, SH$^+$ and HSO$^+$ – are pure X-ray tracers, that is they are more likely to be enhanced by X-rays than by FUV radiation (Sect. 5.1). The ratio CN/HCN increases with increasing X-ray flux whereas the ratios HNC/HCN, HOC$^+$/HCO$^+$ and CO$^+$/HCO$^+$ are not sensitive to the X-ray flux and trace the FUV field as confirmed by observations and models of PDRs by several authors. Whether the CN/HCN ratio is enhanced by an inner FUV field or by X-rays depends mainly on the FUV absorbing inner column density (Sect. 5.2).

5. SH$^+$ and HSO$^+$ are found to be excellent X-ray tracers. They are 100-10000 times more abundant in models with X-rays. Both are more likely to be enhanced by X-rays than by an inner FUV source. In addition, the abundance of SH$^+$ correlates with the X-ray flux. Only four hyperfine lines of SH$^+$ have so far been measured in the laboratory whereas no information is available for HSO$^+$. We would therefore like to encourage laboratories to further investigate these molecules (Sect. 5).

6. CO$_2$ abundances are reduced in the gas-phase through X-ray induced FUV photons. For temperatures $T \lesssim 230$ K, H$_2$O is destroyed by X-rays with luminosities $L_X \gtrsim 10^{30}$ ergs s$^{-1}$ (Sect. 4.1).

7. An enhanced outer FUV field increases the column densities of C, C$^+$ and simple carbon-hydrides like CH and CH$^+$. The influence on the total column density of other species is minor, however (Sect. 4.7).

8. Comparison between observations and models is in general improved with models containing an inner X-ray source. Best-fit models for AFGL 2591 predict an X-ray luminosity $L_X \gtrsim 10^{31}$ ergs s$^{-1}$ with a hard X-ray spectrum $T_X \gtrsim 3 \times 10^7$ K. The ratio of the X-ray luminosity to the total luminosity is found to be $L_X/L_{bol} \approx 10^{-6}$. Best-fit models also confirm the chemical age $t \sim 10^4$ yrs of AFGL 2591 suggested by Doty et al. (2002). The results become independent of time at high X-ray luminosities once $t \gtrsim 10^4$ yrs (Sect. 6.1).

9. Previously proposed X-ray tracers like CN or NO are confirmed by our models. The enhancement of NO, however, is not obvious since it is only prominent in the inner $\approx 500$ AU of the envelope. Small (0.5″–1″) beams are therefore required to observe the enhanced region of NO (Sect. 5.3).

10. Calculated line intensities of $H^{13}CO^+$ and $HCS^+$ show that the enhancement due to X-rays is detectable for these species with single-dish telescopes. Due to the frequency dependence of the critical density, the influence of X-rays is more prominent in higher frequency transitions. High-$J$ lines of $H^{13}CO^+$ can be used to separate the effects of X-ray and cosmic-ray ionization for $L_X \gtrsim 10^{31}$ ergs s$^{-1}$ and $\zeta_{cr} \lesssim 3 \times 10^{-16}$ s$^{-1}$ (Sect. 7).

11. The effects of Compton ionization on the chemistry are studied and compared to results of models without Compton scattering. It is found that the differences are minor and that the effects of Compton ionization can be neglected for column densities $N_H \lesssim 10^{24}$ cm$^{-2}$ (Sect. 2.1).

12. Future instruments like Herschel-HIFI, SOFIA and ALMA are needed to verify our models. In particular, Herschel-HIFI will be able to observe many hydrides whereas the sensitivity and spatial resolution of ALMA is well-suited to measure the size and geometry of the emitting region.

*Acknowledgements.* The authors are grateful to Michiel Hogerheijde and Floris van der Tak for the use of their Monte Carlo code. We thank the referee for the valuable comments. The authors would further like to thank Manuel Güdel, Paolo Grigis and Kevin Briggs for useful discussions. This work was partially supported under grants from The Research Corporation (SDD). Astrochemistry in Leiden is supported by the Netherlands Research School for Astronomy (NOVA) and by a Spinoza grant from the Netherlands Organization for Scientific Research (NWO).

# Online Material

# Appendix A: Initial chemical abundances

**Table A.1.** Initial gas-phase and total abundances.

| Species | $x(i)$ $T > 100$ K | $x(i)$ $T < 100$ K | $x(i)$ total |
|---|---|---|---|
| CO | 1.8E-04[a] | 1.8E-04[a] | |
| $CO_2$ | 1.5E-05[b] | 0.00[f] | |
| $H_2O$ | 7.5E-05[c] | 0.00[f] | |
| $H_2S$ | 8.0E-07[d] | 0.00[f] | |
| $N_2$ | 3.5E-05[e] | 3.5E-05[e] | |
| $CH_4$ | 5.0E-08[e] | 5.0E-08[e] | |
| $C_2H_4$ | 4.0E-08[e] | 4.0E-08[e] | |
| $C_2H_6$ | 5.0E-09[e] | 5.0E-09[e] | |
| $H_2CO$ | 6.0E-08[e] | 0.00[f] | |
| $CH_3OH$ | 5.0E-07[e] | 0.00[f] | |
| He | 8.5E-02[h] | 8.5E-02[h] | 8.5E-02[h] |
| C | 1.0E-08[h] | 1.0E-08[h] | 3.5E-04[i] |
| N | 1.0E-08[h] | 1.0E-08[h] | 1.1E-04[i] |
| O | 0.00[e] | 4.0E-05[g] | 4.5E-04[i] |
| S | 0.00[e] | 3.0E-09[d] | 1.0E-05[i] |
| Si | 1.0E-08[h] | 1.0E-08[h] | 3.5E-05[i] |
| Cl | 1.0E-08[h] | 1.0E-08[h] | 9.3E-08[i] |
| Fe | 1.0E-08[e] | 1.0E-08[e] | 3.2E-05[i] |
| Ne | 1.0E-08[h] | 1.0E-08[h] | 1.4E-04[i] |
| Na | 1.0E-08[h] | 1.0E-08[h] | 2.1E-06[i] |
| Mg | 1.0E-08[h] | 1.0E-08[h] | 4.0E-05[i] |
| Al | 1.0E-08[h] | 1.0E-08[h] | 3.1E-06[i] |
| Ar | 1.0E-08[h] | 1.0E-08[h] | 3.8E-06[i] |
| Ca | 1.0E-08[h] | 1.0E-08[h] | 2.3E-06[i] |
| Cr | 1.0E-08[h] | 1.0E-08[h] | 5.0E-07[i] |
| Ni | 1.0E-08[h] | 1.0E-08[h] | 1.8E-06[i] |

All abundances are relative to total hydrogen. Only the abundances in the 2. and 3. column enter the chemistry. The total abundances $x(i)$ (4. column) are for the attenuation of the X-rays.
[a] van der Tak et al. (1999), [b] Boonman et al. (2003b), [c] Boonman & van Dishoeck (2003), [d] see text, [e] Charnley (1997), [f] assumed to be frozen-out or absent in cold gas-phase, [g] taken to be ~ consistent with Meyer et al. (1998), [h] assumed abundances, [i] Yan (1997).

# Appendix B: Electron recombination and charge transfer reactions

# Appendix C: Influence of $T_X$ and $N_{H,in}$

# Appendix D: Influence of an outer FUV field and cosmic rays

# Appendix E: Comparison to observations

The observation type is listed in Table E.1. Infrared absorption lines are observed along the line of sight toward the central source. The column densities can therefore be compared to the radial column density $N_{radial} = \int n(r) dr$. For centrally condensed envelopes, the column density is dominated by the interior.

Emission lines on the other hand arise from throughout the envelope. The column densities for the emission lines in Table E.1 are calculated from the fractional abundances that were determined through detailed, non local thermodynamic equilibrium (NLTE) radiative transfer modeling (eg., van der Tak et al. 1999). For a density distribution like that of AFGL 2591 (Sect. 2.2), the mass is dominated by the outer part of the envelope and emission measurements often probe the exterior (see also Doty et al. 2002 for a detailed discussion).

**Table B.1.** Electronic recombination reactions.

| Reaction | a | b | Reference |
|---|---|---|---|
| $O^{2+} + e^- \rightarrow O^+ + h\nu$ | 1.9E-11 | -0.65 | AP73 |
| $C^{2+} + e^- \rightarrow C^+ + h\nu$ | 2.2E-11 | -0.65 | AP73 |
| $N^{2+} + e^- \rightarrow N^+ + h\nu$ | 2.1E-11 | -0.64 | AP73 |
| $S^{2+} + e^- \rightarrow S^+ + h\nu$ | 2.0E-11 | -0.69 | MHT96 |
| $Fe^{2+} + e^- \rightarrow Fe^+ + h\nu$ | 2.1E-11 | -0.84 | MHT96 |
| $Si^{2+} + e^- \rightarrow Si^+ + h\nu$ | 1.6E-11 | -0.79 | MHT96 |
| $Mg^{2+} + e^- \rightarrow Mg^+ + h\nu$ | 1.7E-11 | -0.84 | AP73 |
| $Ne^{2+} + e^- \rightarrow Ne^+ + h\nu$ | 1.7E-11 | -0.69 | AP73 |
| $Ne^+ + e^- \rightarrow Ne + h\nu$ | 3.1E-12 | -0.76 | AP73 |

Rate coefficients have the form $k = a(T/300)^b$ and $a$ is in units of cm$^3$ s$^{-1}$. AP73 refers to Aldrovandi & Péquignot (1973). MHT96 refers to Maloney et al. (1996).

**Table B.2.** Charge transfer reactions.

| Reaction | a | b | Reference |
|---|---|---|---|
| $O^{2+} + H \rightarrow O^+ + H^+$ | 7.7E-10 | | Yan 1997 |
| $C^{2+} + H \rightarrow C^+ + H^+$ | 1.0E-12 | | Yan 1997 |
| $N^{2+} + H \rightarrow N^+ + H^+$ | 8.6E-10 | | Yan 1997 |
| $S^{2+} + H \rightarrow S^+ + H^+$ | 1.0E-14 | | Yan 1997 |
| $Fe^{2+} + H \rightarrow Fe^+ + H^+$ | 1.6E-10 | 0.17 | MHT96 |
| $Mg^{2+} + H \rightarrow Mg^+ + H^+$ | 9.0E-14 | | AR85 |
| $Ne^{2+} + H \rightarrow Ne^+ + H^+$ | 1.0E-14 | | AR85 |
| $Si^{2+} + H \rightarrow Si^+ + H^+$ | 1.8E-09 | 0.28 | AR85 |
| $O^{2+} + H_2 \rightarrow O^+ + H_2^+$ | 1.0E-09 | | Yan 1997 |
| $C^{2+} + H_2 \rightarrow C^+ + H_2^+$ | 1.0E-13 | | Yan 1997 |
| $Fe^{2+} + H_2 \rightarrow Fe^+ + H_2^+$ | 1.0E-14 | | Yan 1997 |
| $S^{2+} + H_2 \rightarrow S^+ + H_2^+$ | 1.0E-15 | | Yan 1997 |
| $N^{2+} + H_2 \rightarrow N^+ + H_2^+$ | 3.4E-11 | | Yan 1997 |
| $N^{2+} + He \rightarrow N^+ + He^+$ | 1.3E-10 | | Yan 1997 |
| $C^{2+} + He \rightarrow C^+ + He^+$ | 1.0E-10 | | AR85 |
| $O^{2+} + He \rightarrow O^+ + He^+$ | 7.1E-12 | 0.95 | AR85 |
| $Ne^{2+} + He \rightarrow Ne^+ + He^+$ | 1.0E-14 | | BD80 |

Rate coefficients have the form $k = a(T/300)^b$ and $a$ is in units of cm$^3$ s$^{-1}$. MHT96 refers to Maloney et al. (1996). AR85 refers to Arnnaud & Rothenflug (1985). BD80 refers to Butler & Dalgarno (1980).

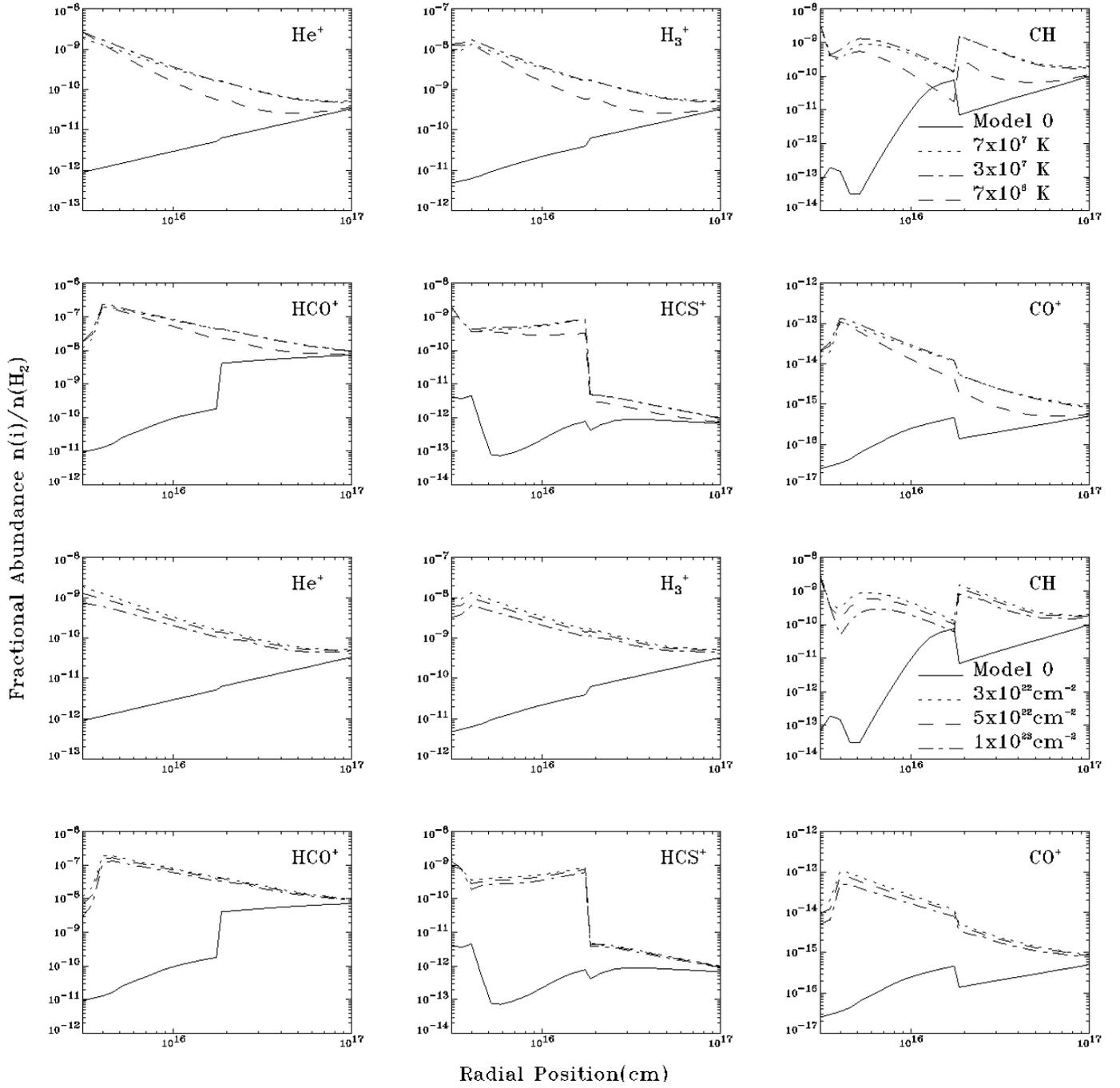

**Fig. C.1.** Depth dependent fractional abundances for different X-ray temperatures $T_X$ with $L_X = 10^{32}$ ergs s$^{-1}$ and $N_{H,in} = 5 \times 10^{22}$ cm$^{-2}$ (upper 6 figures) and for different inner column densities $N_{H,in}$ with $L_X = 10^{32}$ ergs s$^{-1}$ and $T_X = 7 \times 10^7$ K (lower 6 figures).

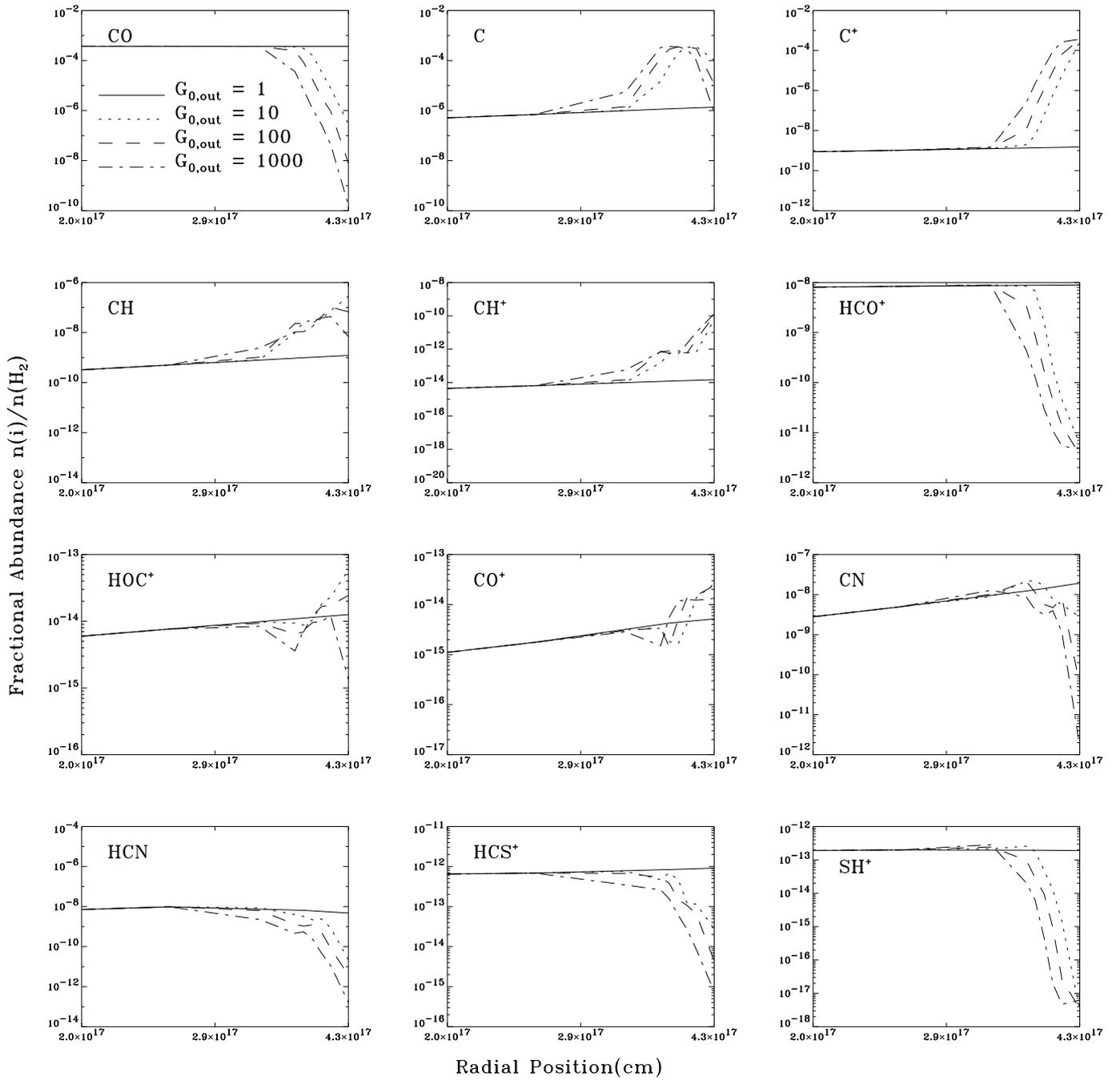

**Fig. D.1.** Depth dependent fractional abundances for different outer FUV fields $G_{0,\mathrm{out}}$ for Model 0.

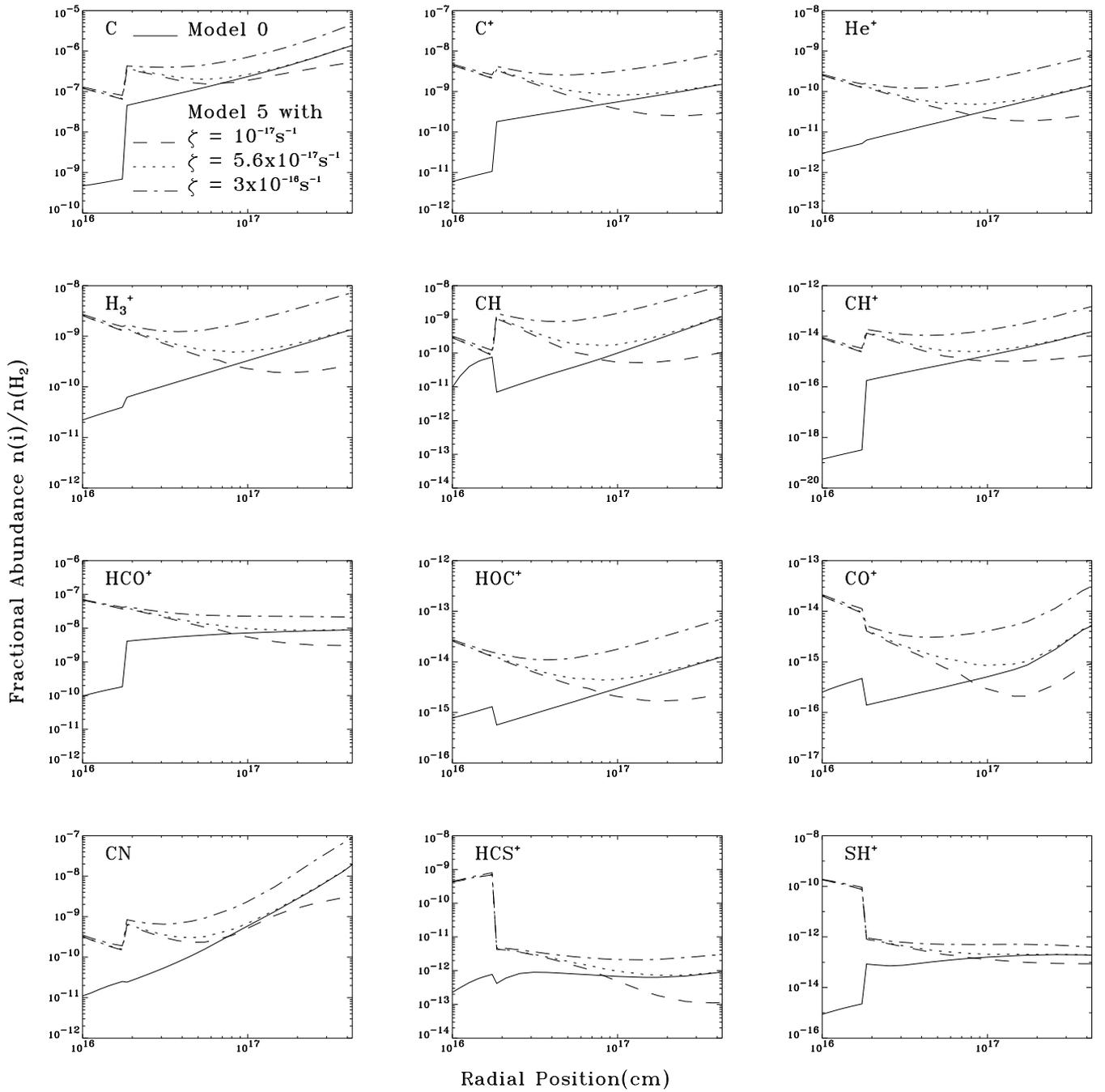

**Fig. D.2.** Depth dependent fractional abundances for different cosmic-ray ionization rates for Model 0 with $\zeta_{cr} = 5.6 \times 10^{-17}$ s$^{-1}$ (solid line), for Model 5 with $\zeta_{cr} = 5.6 \times 10^{-17}$ s$^{-1}$ (dotted line), for Model 5 with $\zeta_{cr} = 10^{-17}$ s$^{-1}$ (dashed line) and for Model 5 with $\zeta_{cr} = 3 \times 10^{-16}$ s$^{-1}$ (dashed-dotted line).

**Table E.1.** Comparison of best-fitting models with observed radial column densities for AFGL 2591.

| Species | Model 6 [cm$^{-2}$] | Observed [cm$^{-2}$] | Data | Reference | $L_X$ [ergs s$^{-1}$] | $T_X$ [K] | $N_{H,in}$ [cm$^{-2}$] | $G_{0,in}$ | Time [yrs] | |
|---|---|---|---|---|---|---|---|---|---|---|
| HCN | 3.8E+16 | 1.2E+16 | submm - JCMT | Bm01 | 3.0E+32 | 5.0E+07 | 4.0E+22 | 0 | 2.4E+04 | |
| HCN | 3.8E+16 | 4.0E+16 | IR - ISO | LvD | 8.0E+31 | 5.0E+07 | 5.0E+22 | 0 | 7.5E+04 | |
| HNC | 4.6E+15 | 9.6E+14 | submm - JCMT | Do | 5.0E+29 | 1.0E+08 | 5.0E+22 | 0 | 5.6E+03 | |
| HC$_3$N | 9.2E+13 | 1.8E+15 | submm - JCMT | Do | 1.0E+30 | 5.0E+07 | 5.0E+22 | 0 | 1.4E+05 | |
| HCO$^+$ | 3.5E+15 | 9.6E+14 | submm - JCMT | vdT99 | 7.0E+30 | 3.0E+07 | 4.0E+22 | 0 | 1.8E+04 | |
| HCS$^+$ | 1.7E+13 | 1.8E+13 | submm - JCMT | vdT03 | 9.0E+31 | 2.0E+07 | 3.0E+22 | 10 | 1.0E+04 | |
| H$_3^+$ | 1.8E+14 | 2-3E+14 | IR - UKIRT | Mc | 2.0E+32 | 3.0E+08 | 7.0E+22 | 0 | 5.6E+03 | |
| H$_2$O | 6.4E+17 | 3.5E+18 | IR - ISO | Bmv | 1.0E+30 | 7.0E+06 | 2.0E+22 | 0 | 4.2E+04 | |
| H$_2$S | 5.3E+12 | 7.3E+14 | submm - JCMT | vdT03 | 5.0E+30 | 3.0E+07 | 3.0E+22 | 0 | 5.6E+03 | |
| H$_2$CO | 2.1E+15 | 3.8E+14 | submm - JCMT | vdT00 | Model 0 | | | 0 | 3.2E+05 | (4.8E+14) |
| H$_2$CS | 2.3E+13 | 2.8E+13 | submm - JCMT | vdT03 | 2.0E+32 | 2.0E+08 | 3.0E+22 | 10 | 1.0E+05 | |
| C$_2$H | 3.8E+14 | 1.9E+14 | submm - JCMT | Do | 6.0E+31 | 1.0E+08 | 5.0E+22 | 0 | 1.0E+05 | |
| C$_2$H$_2$ | 3.3E+14 | 2.0E+16 | IR - ISO | LvD | Model 0 | | | 0 | 5.6E+04 | (1.0E+15) |
| CH$_4$ | 2.6E+16 | 2.5E+17 | IR - UKIRT | Do | 7.0E+32 | 7.0E+07 | 5.0E+22 | 0 | 3.2E+05 | (3.7E+16) |
| CH$_3$OH | 3.4E+13 | 2.8E+15 | submm - JCMT | vdT00 | 5.0E+31 | 1.0E+07 | 1.0E+23 | 0 | 1.3E+04 | |
| CH$_3$CN | 2.6E+14 | 1.8E+15 | submm - JCMT | Do | 1.0E+30 | 5.0E+07 | 1.0E+22 | 0 | 3.2E+05 | (9.5E+14) |
| CO$_2$ | 2.0E+16 | 2.5E+16 | IR - ISO | Bm03 | 3.0E+32 | 3.0E+07 | 3.0E+22 | 0 | 2.4E+05 | |
| CS | 1.9E+15 | 9.6E+14 | submm - JCMT | vdT03 | 6.0E+31 | 1.0E+08 | 3.0E+22 | 10 | 2.4E+05 | |
| CN | 5.6E+14 | 4.5E+15 | submm - JCMT | Do | 5.0E+32 | 5.0E+07 | 5.0E+22 | 0 | 1.8E+04 | |
| OCS | 8.8E+14 | 9.6E+14 | submm - JCMT | vdT03 | 6.0E+31 | 5.0E+07 | 3.0E+22 | 0 | 1.4E+05 | |
| NH$_3$ | 5.7E+15 | 1.9E+15 | cm - Effelsberg | Do | 5.0E+30 | 1.0E+08 | 7.0E+22 | 0 | 1.0E+04 | |
| N$_2$H$^+$ | 4.9E+13 | 4.7E+13 | submm - JCMT | St | 7.0E+31 | 1.0E+08 | 3.0E+22 | 0 | 1.0E+04 | |
| SO | 1.8E+16 | 9.6E+14 | submm - JCMT | vdT03 | Model 1 | | | 10 | 5.6E+04 | (1.6E+15) |
| SO$_2$ | 2.7E+15 | 1.9E+14 | submm - JCMT | vdT03 | 5.0E+32 | 3.0E+07 | 3.0E+22 | 0 | 1.0E+04 | (6.9E+14) |
| NS | 2.4E+12 | 4.7E+11 | submm - JCMT | vdT03 | 6.0E+31 | 3.0E+07 | 3.0E+22 | 10 | 5.6E+03 | (2.0E+12) |

The column density of a model with the parameters $L_X$, $T_X$, $N_{H,in}$, $G_{0,in}$ and time given in columns 6–10 match the observed column density. If no model was found to give a good fit, the best fitted model is presented with the column density given in brackets at the end of the line. References: Bm01 Boonman et al. (2001), Bm03 Boonman et al. (2003b), Bmv Boonman & van Dishoeck (2003), Do Doty et al. (2002), LvD Lahuis & van Dishoeck (2000), Mc McCall et al. (1999), St Stark et al. (1999), vdT99 van der Tak (1999), vdT00 van der Tak (2000), vdT03 van der Tak (2003).